\documentclass[aps,prd,twocolumn,preprintnumbers,nofootinbib,superscriptaddress,amsmath]{revtex4-2}
\usepackage[english]{babel}
\usepackage[utf8]{inputenc}
\usepackage[colorinlistoftodos, color=green!40, prependcaption]{todonotes}
\usepackage[pdftex, pdftitle={Article}, pdfauthor={Author}]{hyperref} 
\usepackage{xcolor}

\usepackage{graphicx}
\usepackage{dcolumn}
\usepackage{bm}

\usepackage{feynmf}

\begin{document}

\preprint{APS/123-QED}

\title{CMB constraints on inflection-point inflation with a pseudo-scalar dark matter}

\author{Jamerson G. Rodrigues}
\email{jamersoncg@gmail.com}
\affiliation{Observatório Nacional, 20921-400, Rio de Janeiro, RJ, Brazil}

\author{Vinícius Oliveira}
\email{vlbo@academico.ufpb.br}
\affiliation{Departamento de Física, Universidade Federal da Paraíba,  58051-970, João Pessoa, PB, Brazil}

\author{Rodrigo von Marttens}
\email{rodrigomarttens@ufba.br}
\affiliation{Instituto de Física, Universidade Federal da Bahia, 40210-340, Salvador, BA, Brazil}%
\affiliation{PPGCosmo, Universidade Federal do Espírito Santo, Vitória-ES, 29075-910, Brazil}

\author{Carlos A. de S. Pires}
\email{cpires@fisica.ufpb.br}
\affiliation{Departamento de Física, Universidade Federal da Paraíba,  58051-970, João Pessoa, PB, Brazil}

\author{Jailson Alcaniz}%
\email{alcaniz@on.br}
\affiliation{Observatório Nacional, 20921-400, Rio de Janeiro, RJ, Brazil}

\date{\today}

\begin{abstract}
In this work, we investigate the physical aspects of the inflection-point inflation scenario and assess its observational viability in light of current Cosmic Microwave Background (CMB) data. The model we consider encapsulates the inflaton with a pseudo-scalar (the dark matter candidate) in a complex neutral scalar singlet. The cosmological constraints on the parameters of inflation derived at a high energy scale are translated to a low energy scale by running these parameters. Ensuring the entire Lagrangian to be invariant under a $Z_3$ symmetry with the adequate transformation of the fields, the imaginary part of the singlet decouples from the other scalars of the model. We then investigate if the observational viability of inflation is also compatible with this pseudo-scalar being the dark matter component.We show that the CMB constraints on the inflationary parameters assure that the pseudo-scalar is stable and provides the correct relic dark matter abundance only when the pseudo-scalar is thermally  produced.
\end{abstract}

\maketitle


\section{\label{sec:intro}Introduction}

The typical energy scales inherent to the early universe make this environment a suitable probe for the physical processes at high energies. Cosmological inflation rises in this context as the paradigmatic explanation for the first moments of the observable universe. Inflation was originally conceived as an early period of accelerating expansion to offer the appropriate initial conditions for the subsequent hot big bang evolution \cite{Guth:1980zm,Linde:1981mu,Albrecht:1982wi,Mukhanov:1981xt,Starobinsky:1982ee}. Such expansion behavior is achieved once the energy content of the cosmos is dominated by an approximately constant component, leading to a nearly de-sitter universe. In the paradigm of slow-roll inflation, this is obtained as the universe follows a flat direction of the inflaton's potential energy. Primordial curvature perturbations are generated once the quantum fluctuations present in the inflaton field are taken into account, seeding the initial condition for the formation of the large-scale structures~\cite{Ross:2014qpa,Alam:2016hwk,Scolnic:2017caz}, as well as the temperature fluctuations present in the Cosmic Microwave Background (CMB) \cite{Ade:2018gkx,Aghanim:2018eyx}.

The present CMB data offers a good test for the numerous constructions of inflationary scenarios \cite{Martin:2013tda}. Some of those are particularly well motivated, according to their capacity to fit the data. This is the case, for example, of the modified gravity scenarios \cite{Starobinsky:1980te,Whitt:1984pd,Salopek:1988qh,Kallosh:2013yoa} and particularly true for the non-minimal Higgs Inflation model \cite{Bezrukov:2007ep}. Nevertheless, these models enhance their capacity to describe observations by adding at least one extra unconstrained parameter, e.g., the non-minimal coupling in Higgs Inflation. By adding complexity to the model, the direct consequence is losing some of its predictability. In addition, the analysis of these inflationary scenarios is based on a set of hypotheses about the evolution of cosmological scales, as pointed out in \cite{Liddle:2003as,Cook:2015vqa,Rodrigues:2023kiz}. This is particularly problematic considering that part of the cosmological expansion is challenging to be observationally constrained, as in the case of the reheating period.

In this work, we propose a realization of the so-called inflection-point inflation (IPI) scenario~\cite{Allahverdi:2006we,Baumann:2007np,Choi:2016eif,Okada:2016ssd,Okada:2017cvy,Bai:2020zil,Ghoshal:2022jeo,Borah:2023csr} and investigate its observational viability in light of the current Planck data \cite{Aghanim:2018eyx}. A near-constant energy density is obtained around a (approximate) saddle point of the inflaton's potential. The conditions according to which this arrangement is possible are obtained following the renormalization group procedure \cite{Sher:1988mj}, fitting the renormalized parameters of the underlying field theory to obtain the appropriate contour conditions leading to the inflationary universe. Moreover, a slow-roll phase is obtained without adding an extra degree of freedom in the model, which increases its predictive aspects. As we will see, the cost is a squeezed tuning of the parameters. Furthermore, assuming that the relevant cosmological scales cross the Hubble horizon very close to the inflection point, we can obtain the predictions for the inflationary observables without further assumptions about the reheating period, reducing the theoretical uncertainties of the model. 

In this study, we employ the Monte-Carlo Markov chain (MCMC) method for precise parameter estimation, focusing on the characterization of the inflationary Lagrangian. Our investigation yields significant insights, allowing us to tightly constrain fundamental aspects such as the amplitude of the inflaton field and its quartic coupling at the pivotal moment of horizon crossing. Furthermore, we also determine the magnitude of the running function denoted as $\beta_\lambda$. These quantities play a crucial role in enhancing our understanding of the underlying low-energy dynamics within the framework of this inflationary scenario.

We also investigate the possible roles of the inflaton field in the phenomenology of particles at low-energies. A particularly interesting possibility follows when inflation is realized in a singlet scalar extension of the standard model (SM). We mean the SM  augmented by one neutral complex scalar singlet ($\phi$). We will see that by imposing the Lagrangian be invariant by a $Z_3$ discrete symmetry with adequate transformation of the fields, we will have an automatic decoupling of the imaginary part of $\phi$ ($I_1$ ) from the other fields, making it a potential DM candidate with the real part ($R_1$) playing the role of the inflaton in the scenario of inflection-point inflation\footnote{For previous studies of DM in the scenario of inflection-point inflation, see \cite{Ghoshal:2022jeo,Okada:2019yne,Ghoshal:2023noe,Ghoshal:2022aoh}}.

This work is organized as follows: In Sec.~\ref{sec:inflation} we implement and develop inflection-point inflation. In Sec.~\ref{sec:model} we present the low energy particle physics model that realizes inflation. In Sec.~\ref{sec:runinng}  we make the running of the parameters responsible for successful inflation from high to low energy scale. In Sec.~\ref{sec:dark} we develop the dark matter sector of the model where we show that the pseudo-scalar that remains in the model is a viable dark matter candidate. In Sec.~\ref{sec:conclusion} we present our conclusions.

\section{\label{sec:inflation}Inflection-Point inflation}

The inflection-point inflation (IPI) is a single-field inflationary scenario induced by a near saddle point in the inflaton potential energy \cite{Allahverdi:2006we,Baumann:2007np,Choi:2016eif,Okada:2016ssd,Okada:2017cvy,Bai:2020zil,Ghoshal:2022jeo,Borah:2023csr}. At leading order, the appropriate Renormalization Group (RG) effective potential resume to
\begin{equation}
    V(\phi) \simeq \frac{\lambda(\phi)}{4}\phi^4, \label{Eq:01}
\end{equation}
where $\phi$ is the inflaton field and $\lambda(\phi)$ is the running quartic coupling evaluated at the background energy scale of the inflaton. For simplicity, the anomalous dimension $\gamma$ of the inflaton is not considered. This general form of the primordial potential allows the analysis of the inflationary dynamics for a wide range of low-energy inspired models. Here we project the primordial potential in the  framework of a scalar singlet dark matter presented below where the natural candidate for inflaton is the real component of the singlet field, which enforces the equivalence between $\phi$ and $R_1$ and the same for $\lambda$ and $\lambda_2$\footnote{Despite the possible confusion that this may cause,  we chose to perform the inflationary analysis considering the notations $\phi$ and $\lambda$ for the inflaton field and its quartic coupling to make explicit use of its generality and avoid cumbersome notation. }(see Eq.~\eqref{eq:pot}). Its behavior is well described by the Taylor expansion for field incursions around the energy scale $M$, 
\begin{equation}
    V(\phi) \simeq V_0 + V_1(\phi - M) + \frac{V_2}{2!}(\phi - M)^2 + \frac{V_3}{3!}(\phi - M)^3\,, \label{Eq:02}
\end{equation}
where terms $V_0$, $V_1$, $V_2$ and $V_3$ are the zeroth, first, second and third order coefficients (derivatives) of the expansion, respectively. In the following we identify $M$ as the amplitude of the inflaton field during the Hubble horizon crossing moment.

From variational calculus, an inflection-point arises whenever the functional form of the potential assumes a null derivative value with changing sign curvature, constraining the coefficients of the Taylor expansion to $V_1=V_2 = 0$. One can also write this coefficients in terms of the renormalized quartic coupling and its variations,
\begin{eqnarray}
    &&\frac{V_0}{M^4} = \frac{1}{4}\lambda(M),\nonumber \\
    &&\frac{V_1}{M^3} = \frac{1}{4}(4\lambda(M) + \beta_{\lambda}(M)),\nonumber \\
    &&\frac{V_2}{M^2} = \frac{1}{4}(12\lambda(M) + 7\beta_{\lambda}(M) + M\beta^\prime_{\lambda}(M)),\nonumber \\
    &&\frac{V_3}{M} = \frac{1}{4}(24\lambda(M) + 26\beta_{\lambda}(M) + 10M\beta^\prime_{\lambda}(M) + M^2\beta^{\prime \prime}_{\lambda}(M)), \nonumber \\ \label{Eq:03}
\end{eqnarray}
where the prime denotes the derivative relative to the inflaton and $\beta_\lambda \equiv \mu \, d\lambda/d\mu$ is the $\beta$-function of the quartic coupling \cite{Okada:2017cvy}. The quantities above are all evaluated at the scale $\phi = M$. In order to recover the conditions for the inflection-point behaviour one must assume\footnote{The third relation is assumed in order to enhance the convergence of the Taylor expansion and avoid the premature end of inflation.},
\begin{eqnarray}
    \beta_{\lambda}(M) &=& -4\lambda(M),\nonumber \\
    \quad M\beta^\prime_{\lambda}(M) &=& 16 \lambda(M),\nonumber \\
    \quad M^2 \beta^{\prime \prime}_{\lambda}(M) &\simeq & -80 \lambda(M). \label{Eq:04}
\end{eqnarray}
Such a specific arrangement for the couplings of the underlying field theory are a promising precursor for the conditions to achieve successful inflation.

\subsection{The Slow-Roll Analysis}

Usually, the analysis carried out for the inflationary period follows from the slow-roll parameters
\begin{eqnarray}
 &\epsilon & = \frac{m^2_{Pl}}{16\pi}\left(\frac{ V^{\prime}}{ V}\right)^2, \quad \quad
 \eta  = \frac{m^2_{Pl}}{8\pi}\left( \frac{V^{\prime \prime}}{V } \right),
 \label{slowparameters}
\end{eqnarray}
where $V$ is the inflationary potential energy, $m_{Pl} = 1.22 \times 10^{19}$ GeV is the Planck mass and the prime denotes the variation with respect to the inflaton field. The slow-roll evolution of the system is accomplished whenever $\epsilon,\eta \ll 1$ and continues until $\epsilon,\eta \simeq 1$.

The study of cosmological perturbations is carried out typically according to the power-law expansion of the primordial power spectrum. For scalar perturbations,
\begin{equation}
    \ln{\frac{P_\mathcal{R}(k)}{P_\mathcal{R}(k_\ast)}} = (n_S-1) \ln{\frac{k}{k_\ast}} + \frac{\alpha_S}{2}\ln{\frac{k}{k_\ast}}^2 + \ldots , \label{Eq:06}
\end{equation}
and similarly for the tensor perturbations, where the ellipsis represents high order in the logarithm and $k_\ast$ is the pivot scale. In particular, the Planck Collaboration has set constraints to the amplitude of scalar perturbations, $P_\mathcal{R}(k_\ast=0.05\,\, \text{Mpc}^{-1}) = 2.101^{+0.031}_{-0.034} \times 10^{-9}$, as well as for the spectral index, $n_S = 0.965 \pm 0.004$ \cite{Aghanim:2018eyx}. Also, an upper limit was obtained for the tensor-to-scalar ratio, $r < 0.06$.

For most of the cosmological scales, the early crossing of the Hubble horizon occurred well inside the slow-roll region ($\epsilon,\eta \ll 1$), where the inflationary observables can be safely associated to the slow-roll parameters
\begin{equation}
    n_S=1-6\epsilon+2\eta \qquad {\rm and} \qquad r=16\epsilon.\label{Eq:07}
\end{equation}
Also, the amplitude of primordial scalar perturbations reads
\begin{equation}
    A_S \equiv P_\mathcal{R}(k_\ast) =\frac{8V}{3m^4_{Pl}\epsilon}. \label{Eq:08}
\end{equation}
All these quantities are evaluated at the inflaton amplitude associated with the horizon crossing moment, $\phi_\ast$. 

Although $\phi_{\ast}$ cannot be directly measured, it can be related to the amount of expansion the universe experienced, from the horizon crossing moment up to the end of inflation. The inflationary number of e-folds is quantified by
\begin{equation}
    N_{\ast} = -\frac{8\pi}{m^2_{Pl}}\int_{\phi_{\ast}}^{\phi_e}\frac{V}{V^\prime}d\phi. \label{Eq:09}
\end{equation}

In this work, we approach in a different direction. First, we assume that the horizon crossing moment for the relevant scales occurred very close to the inflection point of the inflationary potential, $\phi_{\ast} \simeq M$. This means that the amplitude of the inflaton field during horizon crossing is determined by the conditions to accomplish the inflection-point behavior, not by $N_{\ast}$, reducing the theoretical uncertainty of the model \cite{Lesgourgues:2007gp}. Second, we use the data from cosmological perturbations to constrain the coefficients of the inflationary potential given by Eq.~\eqref{Eq:02} and, ultimately, the quantities of the underlying field theory through relations given by Eq.~\eqref{Eq:03}. This will enable us to obtain the predictions for the underlying model based on the high-energy dynamics of the early universe.

To begin with, let's consider the results obtained from the MCMC analysis performed by the Planck Collaboration. In particular, the value obtained for the amplitude of scalar perturbations, $A_S \simeq 2.1 \times 10^{-9}$, allows the relation
\begin{equation}
    V_1 \simeq 2005.16 \left(\frac{\sqrt{8\pi V_0}}{m_{Pl}}\right)^3,\label{Eq:10}
\end{equation}
which brings to the obvious conclusion that the perfect inflection-point behaviour at horizon crossing, $\epsilon = 0$, is not compatible with observations. The underlying question is how far from perfect flatness should the potential be for the model predictions to describe the observed data? 

In order to align the model predictions to observations one shall deviate the horizon crossing moment from the perfect inflection conditions. The most general conjecture for IPI scenario is obtained when the first and second derivative of the inflationary potential deviate from zero,
\begin{equation}
    \beta_{\lambda}(M) \simeq -4\left[\lambda(M)-a\right], \quad  M\beta^\prime_{\lambda}(M) \simeq 16 \left[\lambda(M)-b\right],\label{Eq:16}
\end{equation}
where $a$ and $b$ are purely phenomenological and infinitesimal parameters that quantify the displacement from the perfect inflection-point behaviour. They are tuned in order to obtain the predictions of the inflationary parameters according to the observations. From the observed value for the amplitude of scalar perturbations as Eq.~\eqref{Eq:10}, we have,
\begin{equation}
    a \simeq \frac{2005.16}{M^3}\left(\frac{\sqrt{8\pi V_0}}{m_{Pl}}\right)^3 \label{Eq:12}
\end{equation}
and the set of coefficients of the inflationary potential are obtained according to Eq.~\eqref{Eq:03}. Once with the values for $a$, $b$ and $M$ specified one can compute the predictions of the model to the spectral parameters and compare to the Planck results.

\begin{figure}[t]
\centering
\includegraphics[width=\columnwidth]{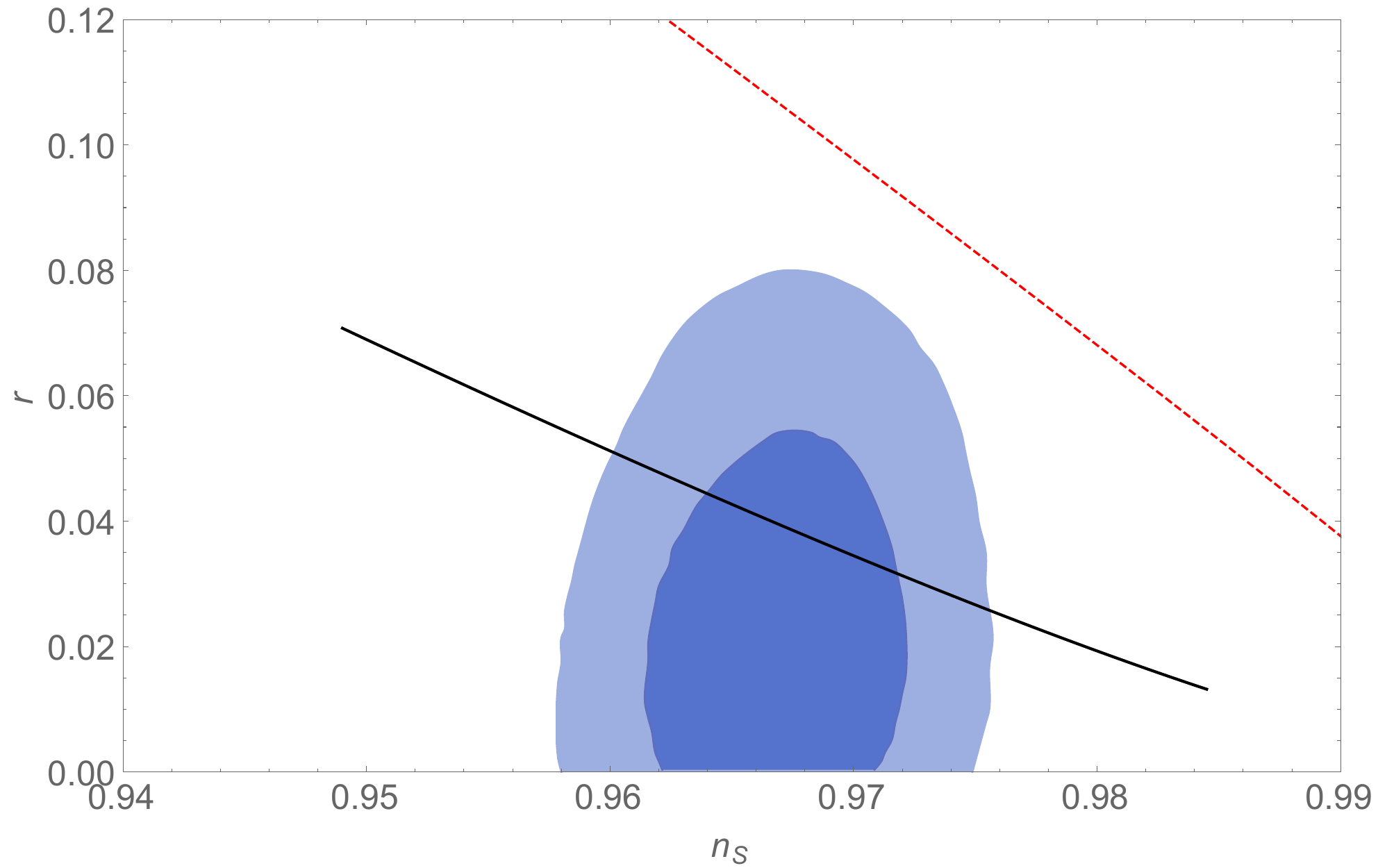}
\caption{$n_S$ vs $r$ for $b=a$ (red dashed) and $b=5a$ (solid black), both with $M = 19.95$ $m_{Pl}$. From the inferior to the superior limit of each curve the parameter $V_0$ assumes the values $1.85 \times 10^{-12} \lesssim V_0/m^4_{Pl} \lesssim 5.91\times 10^{-12}$ (red dashed) and $6.49 \times 10^{-13} \lesssim V_0/m^4_{Pl} \lesssim 3.48\times 10^{-12}$ (solid black). The blue areas show the favoured regions by Planck2018, with $68\%$ and $95\%$ confidence level (Planck $TT,TE,EE + lowE +lensing +  BICEP2/Keck + BAO$ data set) \cite{Aghanim:2018eyx}.}
\label{Fig:ns_r}
\end{figure}

In Fig.~\ref{Fig:ns_r} we present these predictions in the $n_S \times r$ plane. The red dashed curve depicts the predictions for $a=b$, \textit{i.e.}, the simplest conjecture where both inclination and concavity of the potential deviate from zero in the same amount. In this case, the anti-correlation between $r$ and $n_S$ prevents the predictions of the model to overlaps the favoured regions obtained by Planck joint analysis. For $b>7a/4$ the curvature of the inflaton potential presents a negative concavity during horizon crossing, displacing the spectral parameters to the Planck's sweet-spot. The solid black curve shows the pattern of predictions for $b=5a$. As one can note, the predictions converge to the favoured region, particularly for $1.56\times 10^{-12} \lesssim V_0/m^4_{Pl} \lesssim 2.18 \times 10^{-12}$, where the $68\%$ confidence level result is recovered. In both cases, the horizon crossing scale is assumed at $M \simeq 19.95\, m_{Pl}$.

The approach presented above provides an useful tool for understanding the conditions under which the proposed inflationary model could describe the early stages of the universe. Nevertheless, it is convenient to recall that the slow-roll analysis inherits a series of inaccuracies from the slow-roll approximations, which tends to displace the predictions of the model. A more precise alternative consists in constraining the inflationary potential directly from the observational data without resourcing to the slow-roll approximations and the Planck's power-law analysis, as pointed out in \cite{Mortonson:2010er} and demonstrated in \cite{Campista:2017ovq,Rodrigues:2020dod,Rodrigues:2021txa,Santos:2023hhk,dosSantos:2021vis} for non-minimal models of inflation. In the next section we describe the Bayesian selection procedure performed in this work and the results obtained for IPI.

\subsection{Parameter selection} \label{sec:param_selec}

In this section, we delineate the methodology employed to compare the predictions of the IPI model with cosmological data. To achieve this, we make use of the Cosmic Linear Anisotropy Solve Software (CLASS)~\cite{lesgourgues2011cosmic,Diego_Blas_2011}, which enables the numerical solution of the Boltzmann equations governing cosmological perturbations and the subsequent derivation of CMB observables. The inherent simplicity of the CLASS coding style facilitates seamless code modification, thereby affording the flexibility to investigate an extensive number of inflationary potentials, some of which are already encompassed within the original codebase. For the IPI scenario, subtle adjustments are introduced to the primordial module of the code to faithfully incorporate the potential described by Eq.~\eqref{Eq:02}.

For the statistical analysis, we employ the MontePython~\cite{Audren:2012wb,Brinckmann:2018cvx} code, bridging the theoretical predictions from CLASS with observational data through an iterative MCMC process, specifically employing the Metropolis-Hastings method. Regarding the observational data, we utilize data from the Planck 2018 release~\cite{Aghanim:2018eyx}. Specifically, our analysis incorporates the {\sc Plik} high-multipole likelihood to evaluate TT, EE, and TE angular power spectra ($l>30$), the {\sc Commander} likelihood for low-$l$ temperature data exclusively, and the {\sc SimAll} likelihood for low-$l$ EE data spanning from $l=2$ to $29$. Additionally, we incorporate the Planck lensing likelihood, reconstructed for lensing multipoles ranging from $8$ to $400$. Collectively, this amalgamation is referred to as the Planck TT,TE,EE+LowE+lensing likelihood, as denoted in \cite{Aghanim:2018eyx}. We employ wide and flat priors to explore parameter space.

Following the analytical framework established in~\cite{Lesgourgues:2007gp}, we derive the predictions for the primordial power spectrum, specifically focusing on the observable inflation, which encompasses approximately 6-7 e-folds in proximity to the horizon crossing time. This approach diverges somewhat from the conventional methodology, where the slow-roll conditions are typically assumed to hold throughout the entirety of inflation. This chosen perspective harmoniously aligns with the IPI scenario discussed earlier, as it embraces the assumption that the Taylor expansion remains valid for amplitudes close to the horizon crossing scale.

The usual parameters describing the dynamics of the late-time universe were employed, encompassing the physical baryon density $\omega_b$ (multiplied by a factor 100), the physical density of CDM ($\omega_{c}$), the angular acoustic scale $\theta_s$ (multiplied by a factor 100), and the optical depth $\tau$. About the early universe dynamics, we deviate from the conventional power-law treatment of the primordial power spectrum as depicted in Eq.~\eqref{Eq:06}, opting instead to directly compute the initial conditions utilizing the inflationary potential described in Eq.~\eqref{Eq:02} \cite{Mortonson:2010er}. Concerning the initial conditions of cosmological perturbations, we opt for the parameters $R_1 =(V_1/V_0)^2$, $R_2= V_2/V_0$, and $R_3 = (V_3/V_0)(V_1/V_0)$. This choice draws inspiration from the slow-roll expressions inherent in the power-law parametrization of the primordial spectrum as represented in Eq.~\eqref{Eq:06}. Such a selection stems from the understanding that the efficacy of Monte Carlo Markov Chains' convergence is often enhanced when the probed parameters exhibit a near-Gaussian distribution~\cite{Lesgourgues:2007gp}. We maintain $R_0 \equiv A_S \simeq 2.1 \times 10^{-9}$ as a fixed reference point, aligning our investigation of scalar perturbation amplitude values with those inferred by the Planck collaboration~\cite{Aghanim:2018eyx}. Collectively, our comprehensive analysis entails the following set of cosmological parameters:

\begin{equation}
    \left\{ R_1,\,\, R_2,\,\, R_3,\,\, 100\omega_b,\,\, \omega_{c},\,\, 100\theta_s,\,\, \tau_{\rm reio}  \right\}\,.
\end{equation}

\begin{figure*}[t]
\centering
\includegraphics[width=0.8\textwidth]{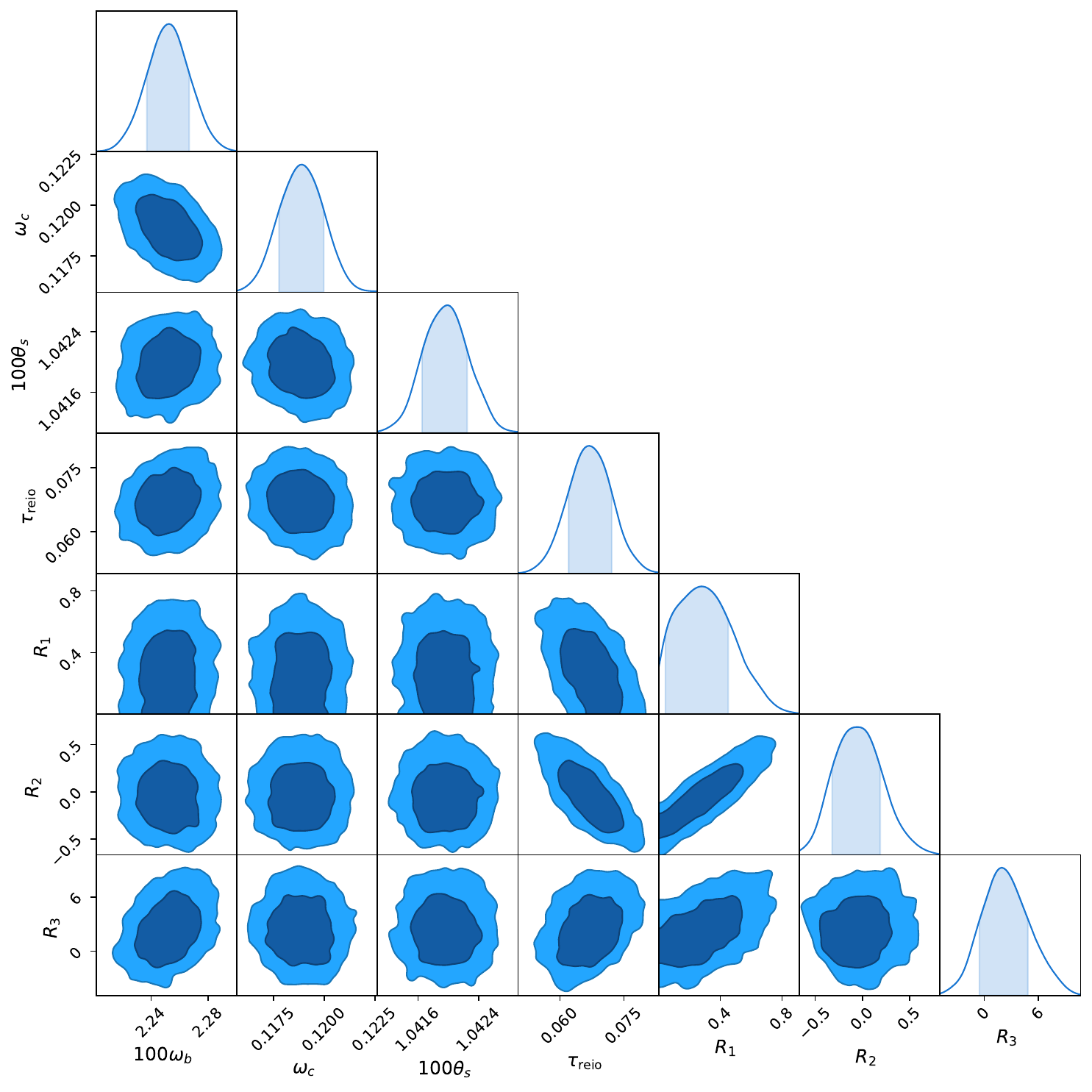}
\caption{Corner plot showcasing the constraints of the cosmological parameters using the Planck TT,TE,EE+lowE+lensing data set~\cite{Aghanim:2018eyx}. Whereas the one-dimensional posteriors highlight in blue the 1$\sigma$ confidence level region, the contours show the 1$\sigma$ and 2$\sigma$ confidence levels. The results for $R_1$, $R_2$ and $R_3$ are presented in units of Planck mass $m_{Pl} = 1$.}
\label{Fig:B1}
\end{figure*}

\subsubsection{Results - Cosmological parameters}
\label{ssec:cosmoparams}

Fig.~\ref{Fig:B1} displays the two-dimensional contour plots and one-dimensional posterior probabilities for the cosmological parameters, with the customary marginalization of the nuisance parameters. The mean values and the $1\sigma$ confidence intervals are also detailed in Tab.~\ref{tab:Cm}. Notably, outcomes for $R_1$, $R_2$, and $R_3$ are presented in units of the Planck mass, denoted as $m_{Pl} = 1$. Convergence behavior adhering to the Gelman-Rubin criterion is satisfactory for all parameters in the chains, with $R-1$ values remaining below $0.01$.

The late-time cosmological parameters are found to be in robust concurrence with those deduced by the Planck collaboration~\cite{Aghanim:2018eyx}. In a reverse exploration akin to the Sec.~\ref{sec:inflation} analysis, one could leverage the $1\sigma$ confidence level constraints for $R_1$, $R_2$, and $R_3$ to deduce the corresponding canonical spectral parameters, following the conventional slow-roll prescription illustrated in Eq.~\eqref{Eq:07}. This approach yields to $n_s = 0.9612 \pm 0.0103$ at the $68\%$ confidence level, which is in good agreement with the Planck analysis considering the scalar fluctuation spectrum in terms of a spectral index $n_s$ and its first derivatives with respect to $\ln k$.

\begin{figure*}
\centering
\includegraphics[width=0.7\textwidth]{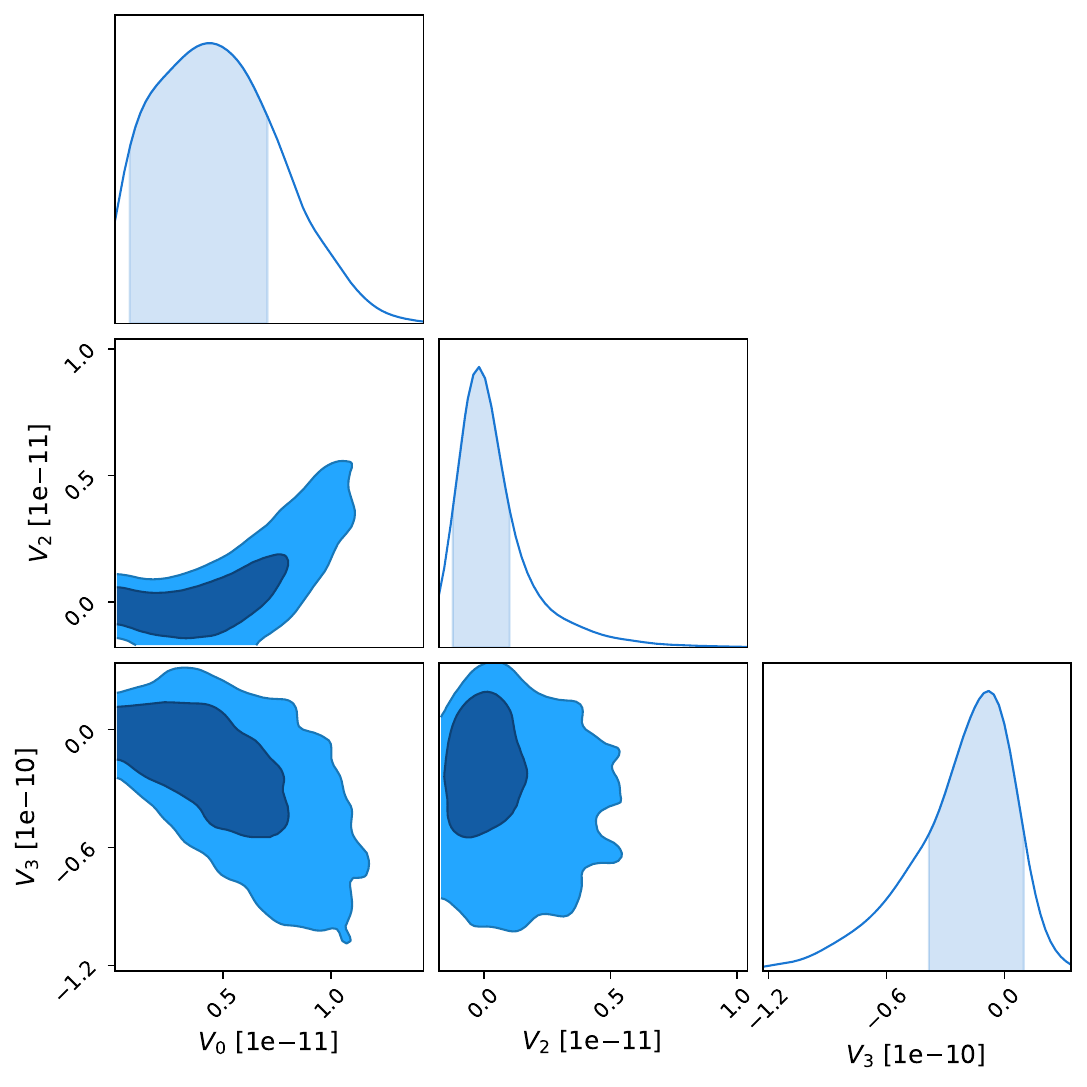}
\caption{Corner plot showcasing the constraints of the derived parameters (Taylor expansion coefficients of the inflationary potential). Whereas the one-dimensional posteriors highlight in blue the 1$\sigma$ confidence level region, the contours show the 1$\sigma$ and 2$\sigma$ confidence levels. The values are presented in units of Planck mass $m_{Pl} = 1$.}
\label{Fig:B2}
\end{figure*}

We extended our analysis to include the coefficients of the inflationary potential as derived parameters. The resultant two-dimensional confidence levels are depicted in Fig.~\ref{Fig:B2}, with corresponding $1\sigma$ intervals thoughtfully compiled within Tab.~\ref{tab:Cm}. As elaborated in Sec.~\ref{sec:inflation}, these parameters directly correlate to the fundamental quantities of the underlying field theory that supports the inflationary model. The energy density associated with inflation remains notably below the Planckian threshold, represented as $V_0 \ll m^4_{Pl}$, thus signifying  favorable stability against potential corrections arising from higher-dimensional Planck-suppressed operators~\cite{Dvali:2020cgt,Burgess:2020nec}.

\begin{table}
\centering
\begin{tabular}{lc}
\hline\hline
\multicolumn{2}{c}{$\quad$ Cosmological parameters $\quad$} \\ \hline
$\omega_b$    & $0.02254^{+0.00013}_{-0.00017}$    \\
$\omega_{c}$   & $0.1189 \pm 0.0011$    \\
$100\theta_s$    & $1.04199^{+0.00026}_{-0.00034}$  \\
$\tau_{\rm reio}$    & $0.0670^{+0.0051}_{-0.0050}$  \\
$R_1$            & $0.29^{+0.16}_{-0.24}$    \\
$R_2$            & $-0.12^{+0.31}_{-0.20}$   \\
$R_3$            & $1.7^{+3.2}_{-2.2}$    \\ \hline\hline
\multicolumn{2}{c}{Derived parameters}      \\ \hline
$10^{12}V_0$     & $4.5^{+2.6}_{-3.8}$      \\
$10^{12}V_2$     & $-0.2^{+1.2}_{-1.0}$      \\
$10^{11}V_3$     & $-0.7^{+1.7}_{-3.1}$     \\ \hline\hline
\end{tabular}
\caption{Constraints on the cosmological parameters within 1$\sigma$ confidence level. The results for $R_1$, $R_2$ and $R_3$ are presented in units of Planck mass $m_{Pl} = 1$, as for $V_0$, $V_2$ and $V_3$.}
\label{tab:Cm}
\end{table}

\begin{figure}
\centering
\includegraphics[width=\columnwidth]{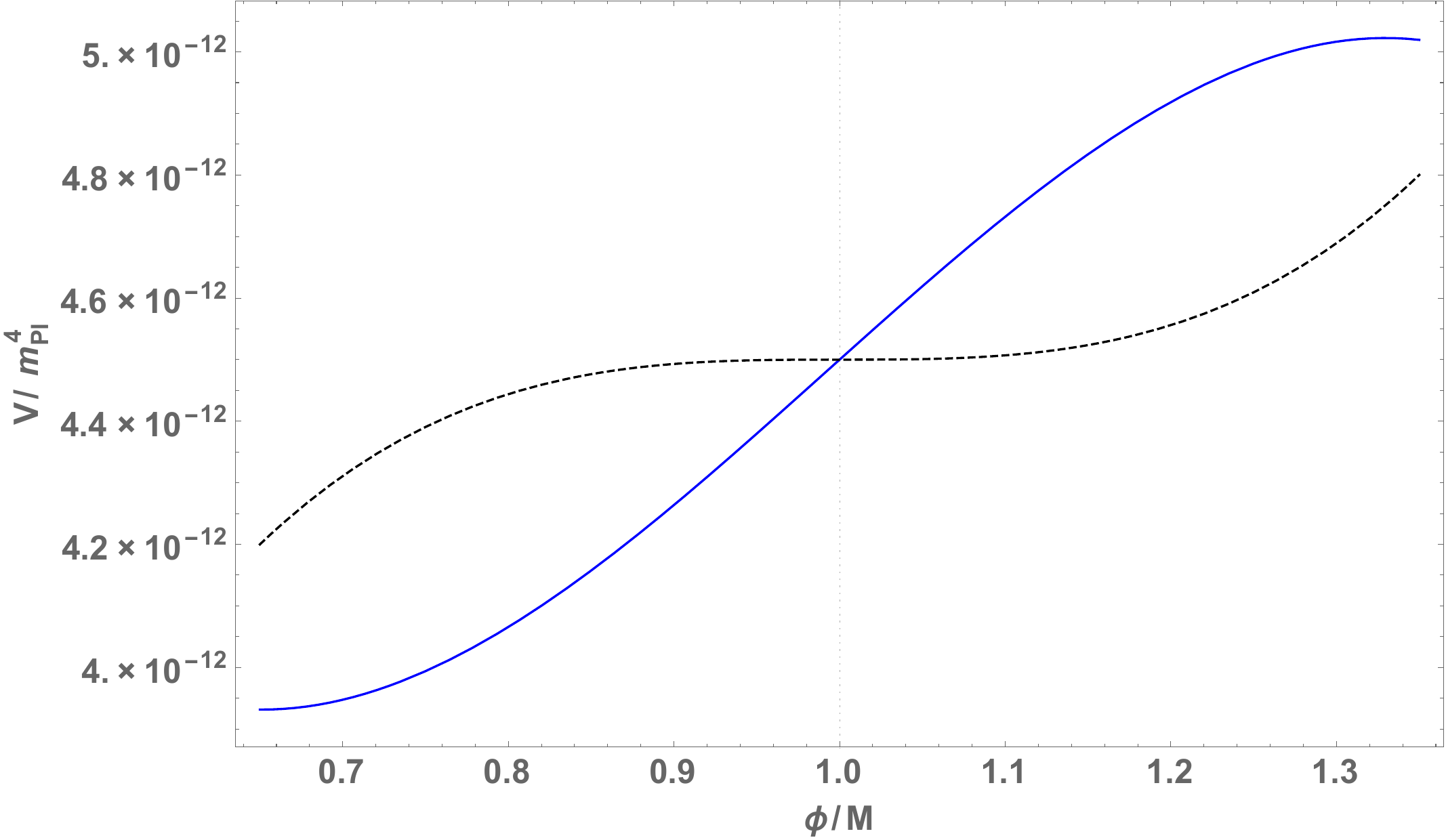}
\caption{Scalar potential for $10^{12}V_0 = 4.5$, $10^{12}V_2 = -0.2$ and $10^{12}V_3=-0.7$ represented in the solid blue line. The black dashed line represents the same potential, but with $10^{12}V_0 = 4.5$, $10^{12}V_1 = 0$, $10^{12}V_2 = 0$ and $10^{12}V_3= 0.7$, while the dashed gray line emphasizes the horizon crossing amplitude $\phi = M$.}
\label{Fig:Pot_V0V2V3}
\end{figure}

For completeness, we also show in Fig.~\ref{Fig:Pot_V0V2V3} the inflationary potential resulting from the MCMC analysis and the definition in ~\eqref{Eq:02}. The solid blue line in the figure is obtained for the central values of the Taylor coefficients in Tab.~\ref{tab:Cm}, while the black dashed line represents the corresponding scenario with exact inflection-point behaviour ($V_1,\, V_2 = 0$).

\subsubsection{Results - Lagrangian parameters} \label{ssec:specific}

{In order to establish a link between the observational results from the CMB and particle physics, constraints on the Lagrangian parameters $\lambda$, $\beta_{\lambda}$ and $M$ can be derived by employing Eqs.~\eqref{Eq:03} and~\eqref{Eq:16}. Following a series of algebraic manipulations, it is straightforward to obtain two set of solutions, given by:}
\begin{eqnarray}
    M^{(\pm)}&=&\Theta_{\pm}V_3^{-1},\nonumber \\
    \lambda^{(\pm)}&=&\frac{2.5\times 10^7 V_0 V_3^4}{\Theta_{\pm}^{4}}\,, \nonumber \\
    \beta_{\lambda}^{(\pm)}&=&\frac{V_0 V_3^3}{\Theta_{\pm}^{4}}\left[-V_3\times 10^8+\right. \nonumber \\ 
    &&\left.\frac{\sqrt{V_0}\left(2.50645\times 10^{11} m_{Pl}^{3/2} V_2-1.00258\times 10^9 \sqrt{\Gamma}\right)}{m_{Pl}^{9/2}}\right]\,. \nonumber \\
    \label{eq:solutions}
\end{eqnarray}
where,
\begin{subequations}
\begin{equation}
\Theta_{\pm} = 250 V_2\pm\frac{\sqrt{\Gamma}}{m_{Pl}^{3/2}} \,, 
\end{equation}
\begin{equation}
\Gamma = 62500 m_{Pl}^3 V_2^2-2.20568\times 10^8 V_0^{3/2} V_3\,. \label{Gamma}
\end{equation}
\end{subequations}

\begin{figure}[t]
\centering
\includegraphics[width=0.8\columnwidth]{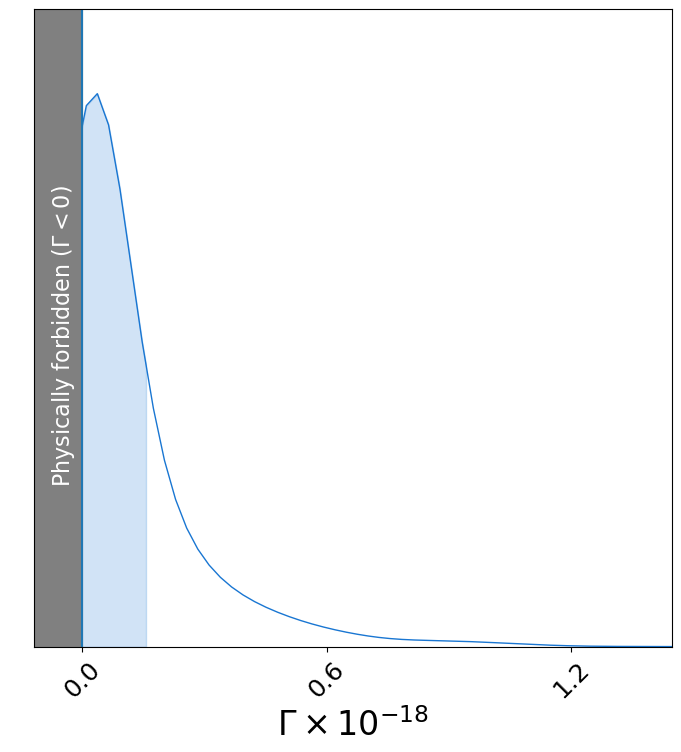}
\caption{Distribution of the term $\Gamma$, as defined in Eq.~\eqref{Gamma}, depicted with the physically prohibited region $\left(\Gamma <0\right)$ shaded in gray. The observational constraints propagated to $\Gamma$ establish an upper bound condition, effectively constraining it to adhere to the limit $\Gamma\leq 1.2\times 10^{-19}$.}
\label{fig:gamma}
\end{figure}

\begin{figure*}
\centering
\includegraphics[width=.32\textwidth]{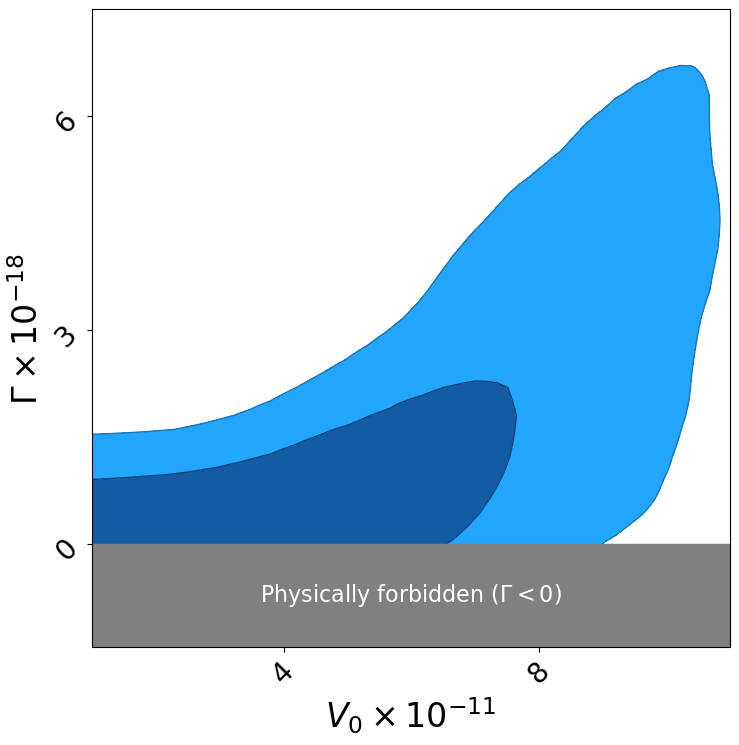}
\includegraphics[width=.32\textwidth]{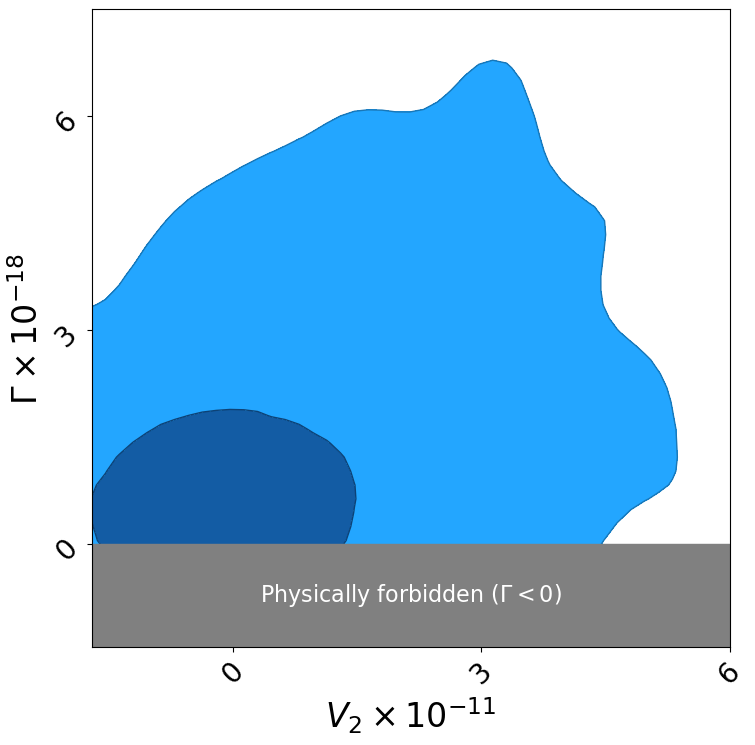}
\includegraphics[width=.32\textwidth]{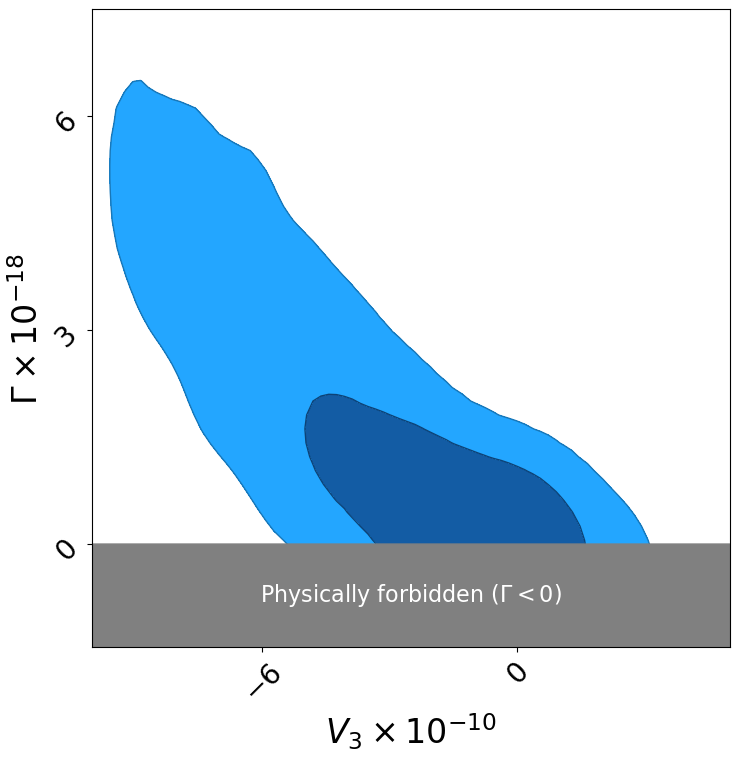}
\caption{Contours of the parameter $\Gamma$ overlaid with the coefficients of the Taylor expansion of the inflationary potential. The gray-shaded region corresponds to $\Gamma<0$, delineating the prohibited domain. Each of the panels showcases a distinct projection: the left panel displays the $\Gamma\times V_0$ plane, the center panel illustrates the $\Gamma\times V_2$ plane, and the right panel exhibits the $\Gamma\times V_3$ plane. They offer a comprehensive insight into the intricate interplay between $\Gamma$ and the inflationary potential's coefficients, emphasizing regions of physical significance and inadmissibility.}
\label{fig:gammaV}
\end{figure*}

An initial consideration regarding the solutions given by Eq.~\eqref{eq:solutions} pertains to their physical validity, restricted to instances where $\Gamma\geq0$. Leveraging the outcomes of the statistical analysis outlined in Sec.~\ref{ssec:cosmoparams}, one can effectively propagate the coefficients of the inflationary potential and subsequently derive the distribution of $\Gamma$. Fig.~\ref{fig:gamma} shows the distribution of $\Gamma$, highlighting the physically prohibited region in gray. Restricting our analysis to the region of physical validity, the resulting distribution of $\Gamma$ imposes an upper limit of $\Gamma\leq 1.2\times 10^{-19}$. Furthermore, Fig.~\ref{fig:gammaV} also illustrates how the $\Gamma\geq0$ condition influences the distributions of the $V_0$, $V_2$, and $V_3$ parameters. To establish a connection between cosmological observations and the scalar extension of the standard model, we will exclusively focus on the region that holds physical significance within this framework.


\begin{figure*}
\centering
\includegraphics[width=0.7\textwidth]{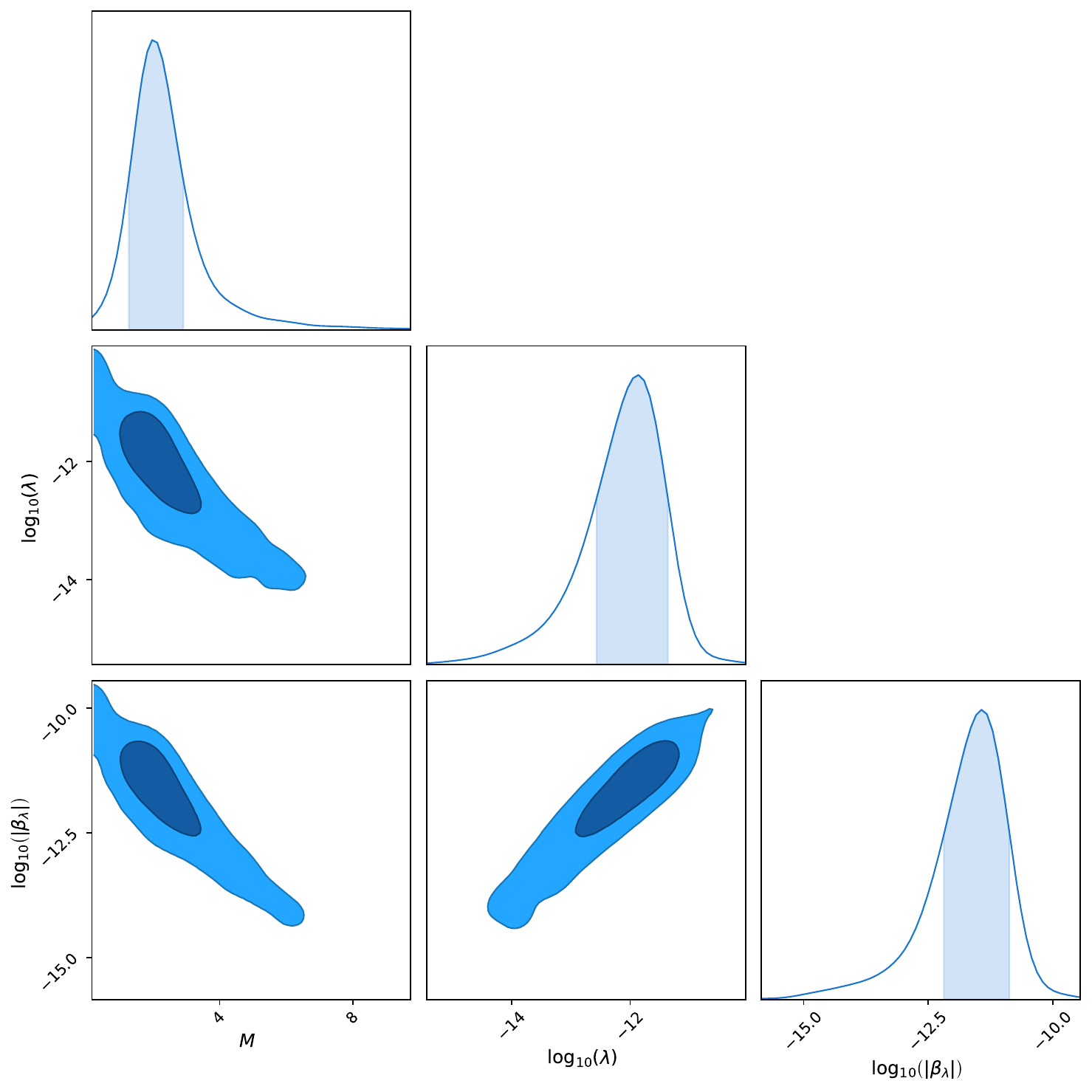}
\caption{Corner plot showcasing the constraints of the Lagrangian parameters. Whereas the one-dimensional posteriors highlight in blue the 1$\sigma$ confidence level region, the contours show the 1$\sigma$ and 2$\sigma$ confidence levels.}
\label{Fig:corner_lagrangian}
\end{figure*}

Both solutions yield highly similar predictions for $\lambda$ and $\beta_{\lambda}$, yet they markedly diverge concerning the parameter $M$. In the context of the (+) solution, a proper propagation of the cosmological parameters leads exclusively to negative values for the energy scale $M$, which is inherently unphysical. In contrast, the (-) solution  yields positive values for $M$. Consequently, from now on, we will exclusively regard the (-) solution as the sole physically viable option, omitting the index (-). The corner plot showing the Lagrangian parameters is presented in Fig.~\ref{Fig:corner_lagrangian}, while the computed mean and corresponding 1$\sigma$ confidence intervals are outlined in Tab.~\ref{tab:lagrangian}.

\begin{table}
\centering
\begin{tabular}{ccc}
\hline\hline
\multicolumn{3}{c}{Lagrangian parameters}                                                            \\ \hline
$M$                    & $\log_{10}\left(\lambda\right)$ & $\log_{10}\left(|\beta_{\lambda}|\right)$ \\
$\qquad1.97^{+0.94}_{-0.70}\qquad$ & $\qquad-11.82^{+0.46}_{-0.75}\qquad$        & $\qquad-11.37^{+0.50}_{-0.82}\qquad$                  \\ \hline\hline
\end{tabular}
\caption{Constraints on the Lagrangian parameters within 1$\sigma$ confidence level.}
\label{tab:lagrangian}
\end{table}

The $68\%$ C.L. limits obtained for the Lagrangian parameters reflect qualitatively some of the well-known results of the high-energy slow-roll scenarios. As usual, the magnitude of the quartic coupling resembles the amplitude of the scalar perturbations, while the inflationary scale $M$ and the running amplitude $\beta_\lambda$ reflect the absence of observable tensorial modes and the quasi-invariance of primordial perturbations with scale. However, the allowed range for the Lagrangian parameters differs in general from the slow-roll analysis performed previously by different authors \cite{Choi:2016eif,Okada:2016ssd,Okada:2017cvy,Ghoshal:2022jeo}. In particular, the value obtained for $M$ in Tab.~\ref{tab:lagrangian} is significantly higher than the superior limit, $M<1.13\,m_{Pl}$, obtained in \cite{Okada:2017cvy}. Also, we have obtained small and negative values for $\beta_\lambda$. Negative contributions to the $\beta$-function of a scalar quartic coupling are generally associated with the presence of fields with fermionic (anti-symmetric) statistics, a condition that will be explored in the next section.

Furthermore, another interesting outcome comes from the combination of~Eqs.~\eqref{Eq:03} and~\eqref{Eq:16}, enabling the determination of $\beta^\prime_{\lambda}$. Following a series of algebraic steps, we obtain,
\begin{eqnarray} \label{beta_prime}
    \beta^\prime_{\lambda}&=&V_3\left[\frac{4 V_2}{M^2}-\frac{56144.5 V_0^{3/2}}{M^3 m_{Pl}^3}+4\times 10^8 V_0 V_3^4 \right. \nonumber \\
    &&\left.\left(250 V_2-\frac{\sqrt{\Gamma}}{m_{Pl}^{3/2}}\right)^{-4}\right]\left(5 V_2-\frac{0.02 \sqrt{\Gamma}}{m_{Pl}^{3/2}}\right)^{-1}\,.
\end{eqnarray}

\begin{figure}[t]
\centering
\includegraphics[width=0.8\columnwidth]{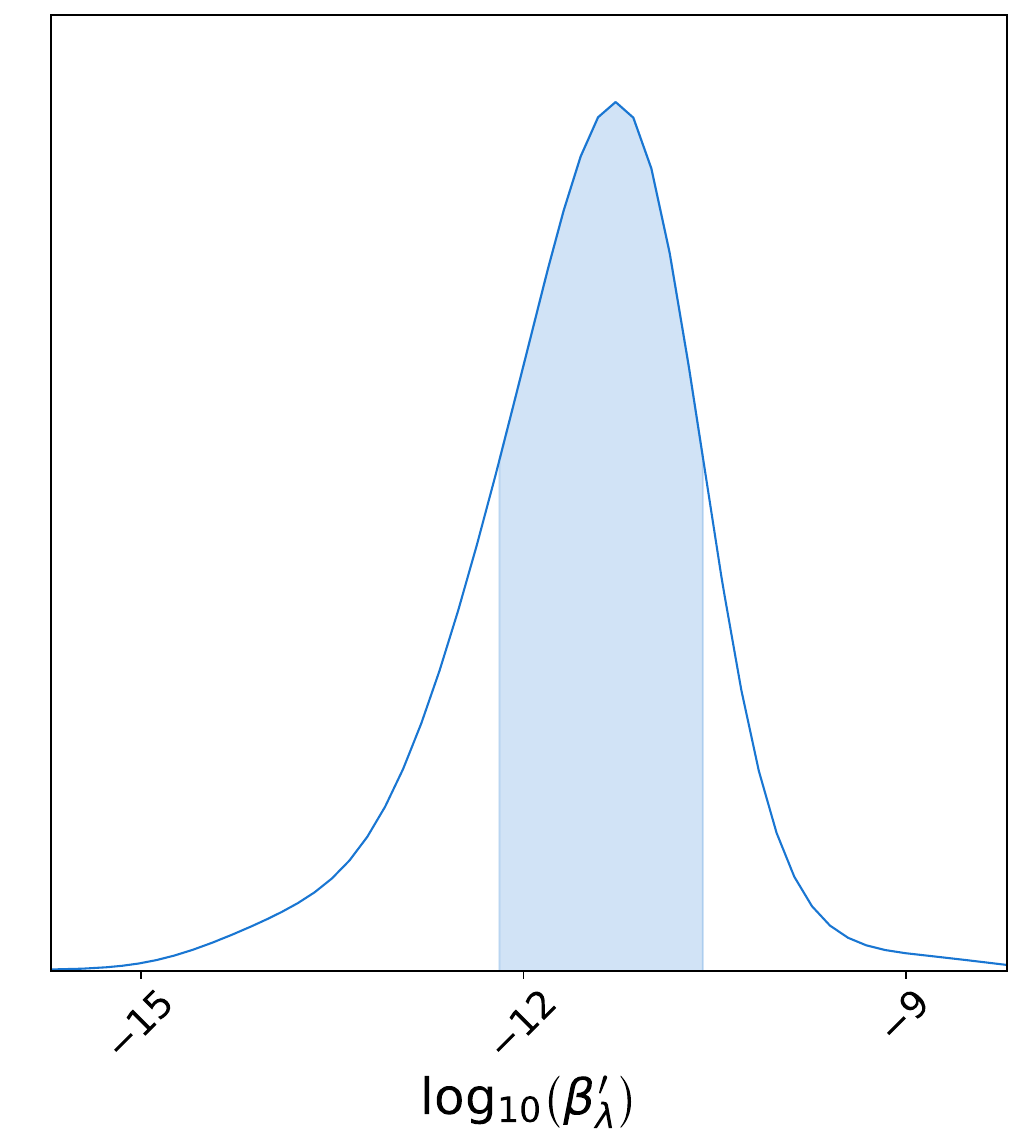}
\caption{Distribution of the term $\beta^\prime_{\lambda}$, as defined in Eq.~\eqref{beta_prime}. The 1$\sigma$ confidence level region is highlighted in the blue region.}
\label{Fig:betaprime}
\end{figure}

Through a proper propagation of the already known parameters, the determination of the $\beta^\prime_{\lambda}$ distribution becomes achievable. Similar to the treatment of $\beta_{\lambda}$, an analysis of the $\log_{10}\left(\beta^\prime_{\lambda}\right)$ distribution is more convenient. Fig.~\ref{Fig:betaprime} shows the outcome of replicating the procedure for propagation of cosmological parameters in Eq.~\eqref{beta_prime}, with $\log_{10}\left(\beta^\prime_{\lambda}\right)=-11.22^{+0.63}_{-0.97}$ within a 1$\sigma$ confidence interval.

\section{The particle physics  Model}
\label{sec:model}

We consider that  the effective potential in Eq.~\eqref{Eq:01} belongs, at low energy scale, to an extension of the SM where the scalar sector is composed by the standard Higgs doublet $\Phi$ and by one neutral complex scalar singlet $\phi$. Moreover, as we will see in the running of the parameters of the model, at least one neutral fermionic singlet $N$ must exist in order to recover the results from inflation. 

The simplest potential involving $\Phi$  and  $\phi$ that do the desired job is obtained by imposing the $Z_3$ discrete symmetry with $(\phi \,, \,N) \rightarrow e^{\frac{2 \pi i}{3}}(\phi\,,\, N)$ and all the other fields transforming trivially by  $Z_3$. In this case the potential allows the following terms 
\begin{eqnarray}
    V(\phi, \Phi) &=& \mu_1^2 \phi^* \phi + \mu_2^2 \Phi^\dagger \Phi + \lambda_1 \left( \Phi^\dagger \Phi \right)^2 + \lambda_2 \left(\phi^* \phi \right)^2 \nonumber \label{Pot:singlet}\\ 
    &+& \lambda_3 \Phi^\dagger \Phi \phi^* \phi -  \frac{M_\phi}{2} \left(\phi^3  + \phi^{\dagger 3}\right)  \,, 
    \label{eq:pot}
\end{eqnarray}
where $\Phi=(\phi^+\,,\, \phi^0)^T$. The last  term  in the potential is important because it generates mass for our dark matter candidate and possesses interactions that contribute to its production. This DM candidate has previously been studied in the context of non-minimal coupling inflation \cite{Cheng:2022hcm}. However, our MCMC analysis for inflection-point inflation scenario yields more stringent constraints on the parameters of the model, additionally imposing the inclusion of a neutral fermionic singlet.

The Yukawa interactions composed by fermions and  scalars that respect $Z_3$ symmetry  are given by
\begin{equation}
    \mathcal{L} =y_d \bar Q \Phi d_{R}+y_u \bar Q \tilde \Phi u_{R} +y_l \bar L \Phi l_{R}+ \frac{1}{2}y_N \bar N^C N \phi  + \text{H.C.}\,,
    \label{eq.yukawa}
\end{equation}
where $Q$, $L$, $d_{R}$, $u_{R}$ and $l_{R}$ are the standard multiplet of fermions. 

In order to achieve our proposal we expand the neutral scalar fields around their respective VEVs,
\begin{eqnarray}
\phi = \frac{v_1 +R_1 + i I_1}{\sqrt{2}}\,, \\
\phi^0 = \frac{v_2 +R_2 +i I_2}{\sqrt{2}}\,,
\end{eqnarray}
and we obtain the set of minimum conditions required by the potential,
\begin{eqnarray}
\mu_1^2  + \lambda_2 v_1^2 + \frac{1}{2} \lambda_3 v_2^2   -
\frac{3 M_\phi v_1 }{2}= 0 \,, \\
\mu_2^2  +  \lambda_1 v_2^2 + \frac{1}{2} \lambda_3  v_1^2 = 0 \,.
\end{eqnarray}

Let us develop the scalar sector according our aim in this  job. For this we consider the CP-even scalars in the basis $(R_1, R_2)$\footnote{From now on, $R_1$ and $R_2$ identify the real components of the fields $\phi$ and $\phi^0$. Care must be taken to avoid confusing them with the cosmological parameters introduced in the previous section.}. Its mass matrix takes the form
\begin{equation}
M_R^2= \left (
\begin{matrix}
2 \lambda_2 v_1^2 - \frac{3 M_\phi v_1}{ 2 }  &  \lambda_3 v_2 v_1  \\
  \lambda_3 v_2 v_1  & 2 \lambda_1 v_2^2 
\end{matrix}
\right) \,.
\end{equation}

We  assume the hierarchy $ v_2 \ll v_1 $. Furthermore inflation will require  $\lambda_2 < \lambda_{3}$.  Then we have that   $R_1$ and $R_2$ mix with each other to form $h$ and $h^{\prime}$ given by
\begin{equation}
\left(
\begin{matrix}
h^{\prime}\\
h
\end{matrix}
\right) = U_R
\left(
\begin{matrix}
R_1\\
R_2
\end{matrix}
\right)
\end{equation}
where
\begin{equation}
 U_R \approx
\left(
\begin{matrix}
-1 &  \epsilon_{R} \\
\epsilon_{R} & 1
\end{matrix}
\right)\,,
\end{equation}
with $\epsilon_{R} = \frac{4 v_2 \lambda_1}{\lambda_3 v_1}$ and 
\begin{equation}
    m_{h^\prime}^2 \simeq 2 v_1^2 \lambda_2 -\frac{3 M_\phi v_1}{\sqrt{2}}\,\,\, \mbox{and} \quad m_{h}^2 \simeq  2 \lambda_1 v_2^2  \,,
\end{equation}
where $v_2 = 245$ GeV, and we took $\lambda_1 = 0.131$ in order to obtain the Standard Higgs mass, $m_h \simeq 125$ GeV.

We have that  $h$ represents the Standard Higgs and $h^{\prime}$ the second Higgs that will play the role of the inflaton. It's crucial to emphasize that $h^\prime$ can be lighter than the Standard Higgs. Furthermore, the aforementioned approximation for the mass equation holds true only for $m_{h^\prime} > 1$ GeV, which falls within our scope of interest.

Note that the second term in the right-hand side of $m_{h^{\prime}}^2$ contribute negatively, which impose the following constraint on $M_\phi$
 \begin{equation}
     M_\phi < \frac{2}{3} \sqrt{2} v_1 \lambda_2\,,
 \end{equation}
 to obtain $m^2_{h^{\prime}}$ positive.

In the CP-odd sector,  we have that  $I_2$ is the Goldstone eaten by $Z^0$ while  $A = I_1$ decouples from $I_2$  and gains the following mass term due to the trilinear term in the potential,
\begin{equation}
    m_{A}^2 = \frac{9 M_\phi v_1}{\sqrt{2}}  \,.
\end{equation}
Concerning the charged scalars, $\phi^{\pm}$, they are  Goldstones eaten by $W^{\pm}$. This is the general aspect of the background  model that leads to inflation  at a high energy scale. In what follows, we incorporate the constraints on $\lambda$ and $\beta_\lambda$ given in Tab.~\ref{tab:lagrangian} on the couplings of the model above in order to see if they are compatible with the pseudo-scalar $A$ playing the role of the dark matter of the universe. 

We remark that the conditions for $A$ to be a dark matter candidate are that it be stable and furnishes the correct abundance of dark matter in the universe. The first is straightforward for $A$ lighter than $N$ which is guaranteed by the $Z_3$ symmetry that is preserved in the Yukawa interaction in Eq. (\ref{eq.yukawa}) while in the potential, after spontaneous symmetry breaking, the stability is guaranteed by  a $Z_2$ symmetry with $A \to -A$\footnote{The potential in Eq. (\ref{eq:pot}) is initialing built invariant by $Z_3$. However here $Z_3$ will be broken spontaneously to a $Z_2$ (this is so because $A$ is a pseudo-scalar and we are assuming CP-invariance). The domain wall that apparently would be generated by the spontaneous breaking of $Z_3$ \cite{Zeldovich:1974uw}  may be evaded by many ways as for example by terms that break explicitly $Z_3$. Such terms may be generated either by gravity \cite{Rai:1992xw}, that does not preserve discrete symmetries, or by effective higher dimension terms in the potential\cite{Larsson:1996sp,Preskill:1991kd}. }.  In what concern the abundance, we will address it in the next section. For while, it is important to highlight that $N$ can be also a DM candidate, since it is stable.

\section{Running of the parameters}\label{sec:runinng}

In Tab.~\ref{tab:lagrangian} we displayed the values of  $\lambda$ and  $\beta_\lambda$ demanded for successful inflation. These parameters correspond to $\lambda_2$ and $\beta_{\lambda_2}$  in the potential in Eq.~\eqref{Pot:singlet}. Then, the evolution of  Renormalization Group Equations (RGEs) of the couplings will provide some constraints on $\lambda_2$, $\lambda_3$  and $y_N$ at low energy scale. The significance of these constraints lies in their potential to determine the viability of $A$ as a plausible dark matter candidate. 

We use SARAH \cite{Staub:2008uz} to compute the RGEs  of the model  at one-loop level, 
\begin{eqnarray} 
    \beta_{\lambda_1} &=& \frac{27}{100}g_1^4 + \frac{9}{4} g_2^4 + \frac{9 }{10} g_1^2 \left( g_2^2 - 2 \lambda_1 \right) - \frac{9}{5} g_2^2 \lambda_1 + 12 \lambda_1^2 \nonumber \\
    &&+ 12 \lambda_1 y_t^{2} - 6 y_t^4  + \lambda_3^2\,, \nonumber  \\  
    \beta_{\lambda_2} &=& 20 \lambda_2^2 + 2 \lambda_3^2  + 2 \lambda_2 y_N^2 - y_N^4\,, \nonumber \\
    \beta_{\lambda_3} &=&  -\frac{9}{10} g_1^2 \lambda_3 - \frac{9}{2} g_2^2 \lambda_3 + 6 \lambda_1 \lambda_3 + 8 \lambda_2 \lambda_3 + 4 \lambda_3^2  \nonumber \\
    &&+ \lambda_3 \left( 6 y_b^2 + 2 y_\tau^2 + y_N^2 + 6 y_t^2 \right)\,, \nonumber \\
    \beta_{y_t} &=& y_t \left( -\frac{17}{20} g_1^2 - \frac{9}{4}g_2^2 - 8 g_3^2 + 3 y_b^2 + y_\tau^2 + 3 y_t^2 \right) \nonumber \\
    &&- \frac{3}{2} \left( y_t y_b^2 - y_t^3  \right)\,, \nonumber \\ \label{eq:run}
    \beta_{y_b} &=& \frac{1}{4} \Big( - \left( y_b \left( g_1^2 + 9 g_2^2 + 32 g_3^2 -12  y_b^2 - 4  y_\tau^2 - 12 y_t^2 \right) \right) \nonumber \\
    &&+ 6 \Big(  y_b^3 -  y_b y_t^2 \Big) \Big)\,,  \\
    \beta_{y_\tau} &=& y_\tau \left( - \frac{9}{4} \left(g_1^2 + g_2^2 \right) + 4 y_\tau^2 +  3 y_t^2 \right) + \frac{3}{2}  y_\tau^3  \,, \nonumber \\
    \beta_{y_N} &=& \frac{3}{2} (y_N)^3\,,\nonumber  \\
    \beta_{g_1} &=& \frac{47}{10} g_1^3\,, \nonumber \\
    \beta_{g_2} &=& - \frac{5}{2} g_2^3\,, \nonumber \\
    \beta_{g_3} &=& - 7 g_3^3 \,,\nonumber 
\end{eqnarray}
where $y_N$ represents the Yukawa coupling of the neutral singlet $N$, $y_t$ represents the Yukawa coupling of top quark, $y_\tau$ represents the Yukawa coupling of tau, $y_b$ represents the Yukawa coupling of bottom, and $g_1$, $g_2$ and $g_3$ are the gauge couplings of the standard gauge group U$(1)$, SU$(2)$, and SU$(3)$, respectively. The value at high energy of $\lambda_2$ and $\beta_{\lambda_2}$ are shown in Tab.~\ref{tab:lagrangian}. Observe that successful inflation requires $\beta_{\lambda_2}<0$.  As we can see in Eq.~\eqref{eq:run} the unique negative contribution to $\beta_{\lambda_2}$ comes from Yukawa coupling of the neutral fermionic  singlet  $N$. This is the reason for the existence of $N$. 

\begin{figure}[t]
    \centering
    \includegraphics[width=\columnwidth]{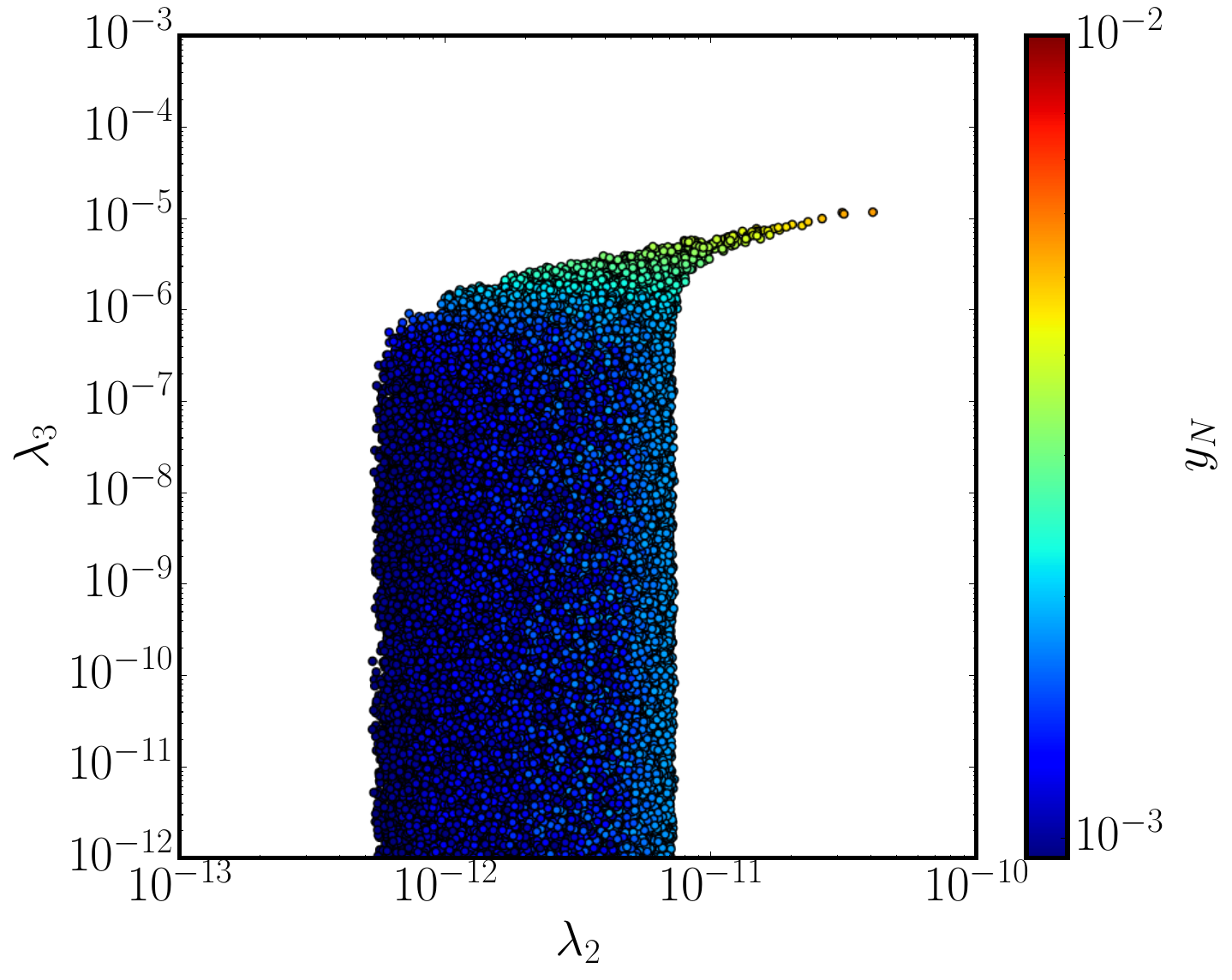}   
    \caption{Viable parameter space in the plane $(\lambda_2, \lambda_3)$, where the color of the points represents the variation of $y_N$, for the running of parameters at one-loop level at low energy scale that recover the values of these parameters at high energy scalar as  required by inflation. }
  \label{fig:Running}
\end{figure}

Now, we can compute the running of the parameters in Eq.~\eqref{eq:run} as a function of energy scale $M$ up to $M = 1.97^{+0.94}_{-0.70} m_{Pl}$. First, we solve the system in Eq.~\eqref{eq:run}. Then, we vary stochastically the initial values of the free parameters ($\lambda_2$, $\lambda_3$, and $y_N$) from low energy scale to high energy with the aim of  obtaining the values required by inflation at energy scale $M$, as shown in Tab.~\ref{tab:lagrangian}. The initial values of the other parameters are shown in Tab.~\ref{tab:initial_value}. Our results are shown in Fig.~\ref{fig:Running} where we have $8\times 10^{-4} \lesssim y_N\lesssim 3\times 10^{-3}$ , which recover the negative value of $\beta_{\lambda_2}$, as required by inflation analysis, and $\lambda_2$ lying in the range  $ 5.3 \times 10^{-13} \lesssim \lambda_2 \lesssim 1.3 \times 10^{-10}$.

\begin{table}
\tiny
\centering
\begin{tabular}{ccccccc}
\hline\hline
\multicolumn{7}{c}{Initial Value}                                              \\ \hline
$\lambda_1$ & $y_t$    & $y_b$     & $y_\tau$  & $g_1$    & $g_2$    & $g_3$   \\ 
$0.258196$    & $0.996819$ & $0.0241282$ & $0.0102566$ & $0.343518$ & $0.641405$ & $1.17571$ \\ \hline\hline
\end{tabular}
\caption{Initial value of parameters.}
\label{tab:initial_value}
\end{table}

Within our model, two potential candidates emerge for playing the role of dark matter: the pseudoscalar $A$ and the fermion $N$. However, the constraint on the parameter $y_N$ coupled with the values of  $\lambda_{2,3}$, results into two important regimes for $m_N$: $m_N > T_R$, and $m_N < T_R$. Where $T_R$ is the reheating temperature. When $m_N > T_R$ the number density of $N$ is Boltzmann suppressed, and then $N$ cannot thermalize. Additionally, the non-thermal $N$ cannot be produced through inflaton decays, since $2m_N \gg m_{h^\prime}$. Furthermore, for  $m_N < T_R$ we could verify that due the high value of $m_N$ and the tiny values of couplings imposed by inflation it also does not thermalize. However, it still can be produced through SM particles in the thermal bath, with enough energy to annihilate in it, but its abundance is very suppressed. In light of this, we only compute the contribution of pseudoscalar $A$ as our dark matter candidate and investigate its  production modes and obtain its abundance.

\section{Dark Matter} \label{sec:dark}
In this section, we assume that $A$ plays the role of the dark matter of the universe. As we will show, the allowed range of values of the parameters $\lambda_{2,3}$ allowed by successful inflation will infer in the production mode of $A$. It is worth mentioning  that  $y_N \sim \mathcal{O}\left( 10^{-3} \right)$ leads to $N$  heavier than $A$, which guarantees $A's$ stability. Additionally, we also observe that the slight allowed variation of $\lambda_2$ in the Fig.~\ref{fig:Running} does not  drastically alter our final DM result; then, we will fix $\lambda_2 = 10^{-12}$.

\begin{figure}
\centering
\unitlength = 1mm
\begin{fmffile}{process1}
\begin{fmfgraph*}(45,15)
\fmfleft{i1,i2}
\fmfright{o1,o2}
\fmflabel{$A$}{i1}
\fmflabel{$A$}{i2}
\fmflabel{SM}{o1}
\fmflabel{SM}{o2}
\fmf{dashes}{i1,v1,i2}
\fmf{plain}{o1,v2,o2}
\fmf{dashes,label=$h,,h^\prime$}{v1,v2}
\end{fmfgraph*}
\end{fmffile}
\medskip
\caption{ Process of $A$ with SM particles through Higgs ($h$) and inflaton ($h^{\prime}$).}
\label{Fig:process}
\end{figure}
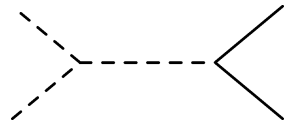

In the early universe, the SM particles are in thermal equilibrium with each other composing, in this way, the thermal bath. Our dark matter candidate $A$ interacts with the SM particles via Higgs $h$ and inflaton $h^{\prime}$ as shown in Fig.~\ref{Fig:process}, with the vertices  
\begin{eqnarray}\label{eq:coupling}
    g_{h A A} \simeq \frac{3  \lambda_1v_2 M_\phi}{v_1 \lambda_3} + \frac{2 v_2 \lambda_1 \lambda_2}{ \lambda_3} + \frac{v_2 \lambda_3}{2}\,, \nonumber \\
    g_{h^{\prime} A A} \simeq \frac{v_2^2 \lambda_1}{v_1} - \frac{M_\phi}{2} - v_1 \lambda_2 \,,
\end{eqnarray}
{where the interaction mediated by Higgs is always the dominant, for our benchmark value of parameters imposed by inflation.} When the rate of the interactions $\Gamma \equiv n \langle  \sigma v \rangle$ -- where $n$ is the number density and $\langle  \sigma v \rangle$ is the thermal averaged cross-section -- is bigger than the Hubble rate $H$ for any temperature $T$ in the early Universe ($\Gamma>H$), $A$ is in thermal equilibrium with SM particles. In this case, $A$ is produced through freeze-out, whereas for $\Gamma <H$, $A$ is produced through freeze-in.

Let us compute the thermally averaged cross-section for the processes shown in Fig.~\ref{Fig:process}. In general we have that 
\begin{eqnarray}\label{thermally_avg}
    \langle \sigma v \rangle \equiv \frac{1}{(n_{A}^{eq}(T))^2} \frac{\cal{S}}{32 \left(2 \pi \right)^6} T \int ds \frac{\sqrt{\lambda \left(s, m_{A}^2,m_{A}^2\right)}}{s} \nonumber \\
    \frac{\sqrt{\lambda \left(s, m_{SM}^2,m_{SM}^2\right)}}{\sqrt{s}} K_1 \left(\frac{\sqrt{s}}{T} \right) \int d \Omega |\mathcal{M}|^2 \,,
\end{eqnarray}
where $\cal{S}$ is the symmetrization factor, $s$ is the Mandelstam variable, $\lambda \left( x, y, z \right)$ is the Källen function, $K_i$ is the modified Bessel function of the second kind of order $i$, $\Omega$ is the solid angle between initial and final states in the center of mass frame, and $|\mathcal{M}|^2$ the (not averaged) squared amplitude of the process. To obtain the thermally averaged cross section we implemented the model in Feynrules and export it to CalcHep to obtain the squared amplitude of the processes. After that we compute the $\langle  \sigma v \rangle$ through CUBA \cite{Hahn:2004fe}, a library of numerical integration in C++.

\begin{figure*}[t]
\centering
\includegraphics[width =0.8\textwidth]{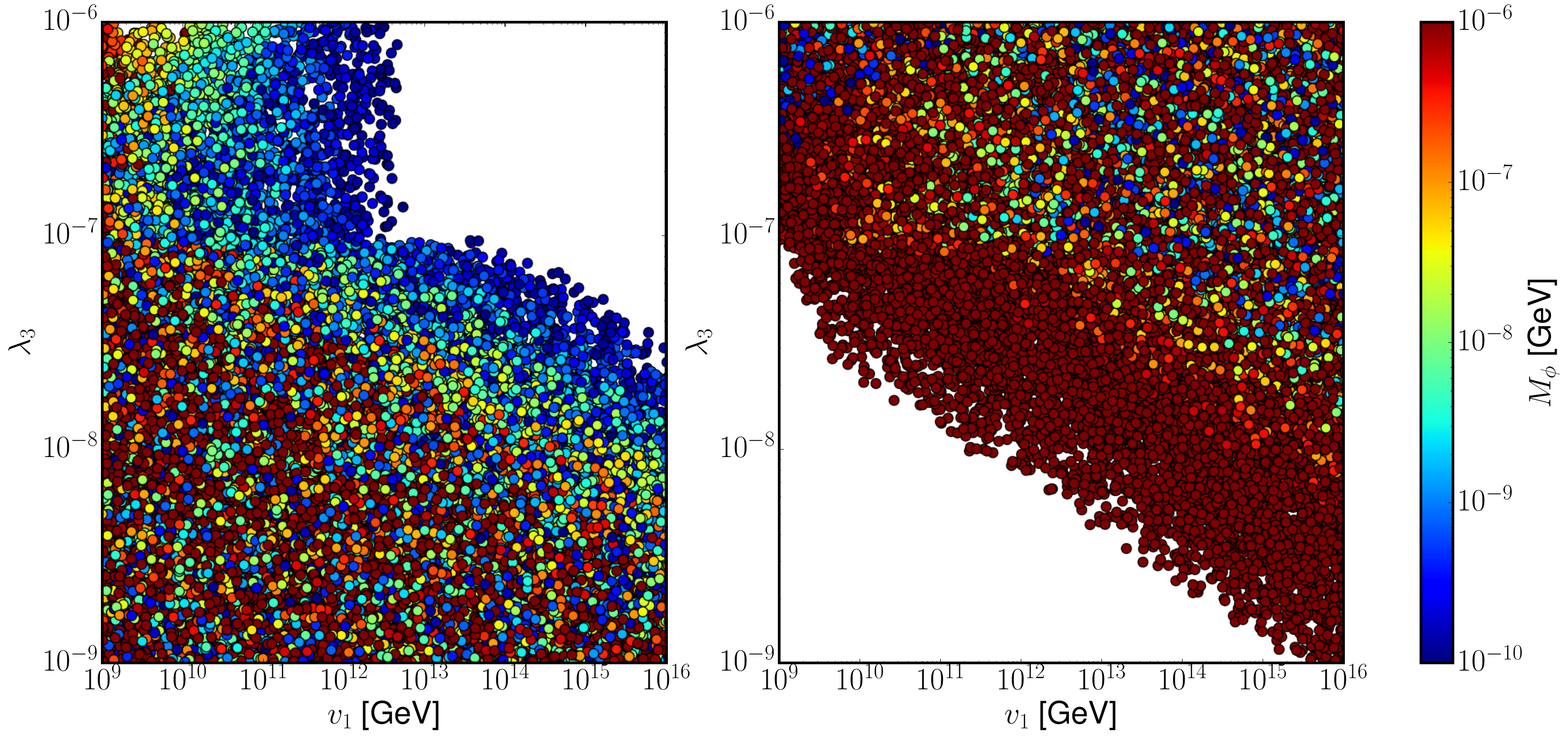}
\caption{Viable parameter space in the plane $\left(v_1, \lambda_3 \right)$ for freeze-out (left panel) and freeze-in (right panel). In both cases we varied $\lambda_3$, $v_1$ and $M_\phi$ stochastically while computing the ratio $\Gamma/H$ for $A$. }
\label{Fig:freezeout_in}
\end{figure*}

It is important to note that four free parameters contribute to $\langle \sigma v \rangle$ for the processes shown in Fig.~\ref{Fig:process}, namely $v_1$, $\lambda_{2,3}$ and $M_\phi$.  After computing $\langle \sigma v \rangle$ varying stochastically all these four parameters, we also compute the ratio $\Gamma/H$ and obtain the parameter of space associated with the thermal and non-thermal production of $A$. The result is shown in Fig.~\ref{Fig:freezeout_in}. The left panel represents the region of parameter space in the plane $\left(v_1, \lambda_3 \right)$ that implies  freeze-out ($\Gamma/H \gg 1$), while the right panel implies freeze-in ($\Gamma/H \ll 1$). In both panels, the color of the points represents the variation of $M_\phi$. It is important to highlight that the points of Fig.~\ref{Fig:freezeout_in} are in agreement with inflation, as shown in Fig.~\ref{fig:Running}.

\begin{figure*}
\centering
\includegraphics[width=0.7\textwidth]{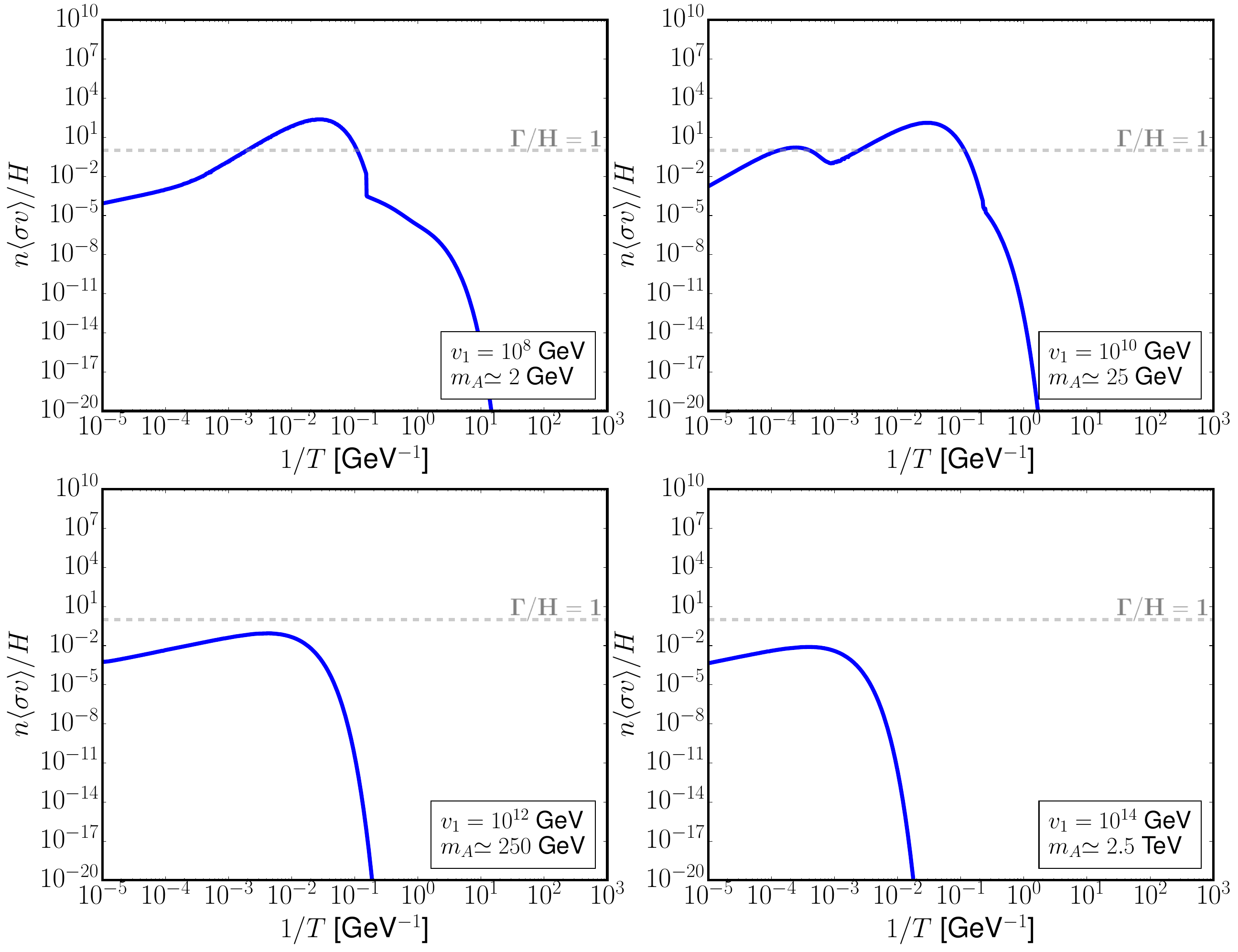}
\caption{Ratio between the interaction rates $\Gamma$ and the Hubble rate $H$ for $A$ varying $v_1$, and fixing $\lambda_2 = 10^{-12}$, $\lambda_3 = 10^{-6}$, and $M_\phi = 10^{-8}$ GeV.}
\label{Fig:vs_thermal}
\end{figure*}

In order to understand more deeply what happens with the ratio $\Gamma/H$ when we vary the parameters, we compute $\Gamma/H$ varying each of these parameters separately. In Fig.~\ref{Fig:vs_thermal}, we show the results for $\Gamma/H$ varying  $v_1$  and  all the other parameters fixed. In the first upper panel, we show the case $m_{h} \sim m_{h^\prime}$, where the peak  corresponds to the $h$ and $h^\prime$ resonance, and it is important to recall $m_{h^\prime}$ increases with $v_1$. Then, when $v_1$ increases, the resonance of $h^\prime$  translates to the left. This case is related to a period of high temperature. Note also that the higher $v_1$ is,  the tinier   $g_{hAA}$ gets.  Consequently, $A$ decouples from the thermal bath, as we seen in the down left panel. This is in complete agreement with the results displayed in Fig.~\eqref{Fig:freezeout_in}.

\begin{figure*}
\centering
\includegraphics[width=0.7\textwidth]{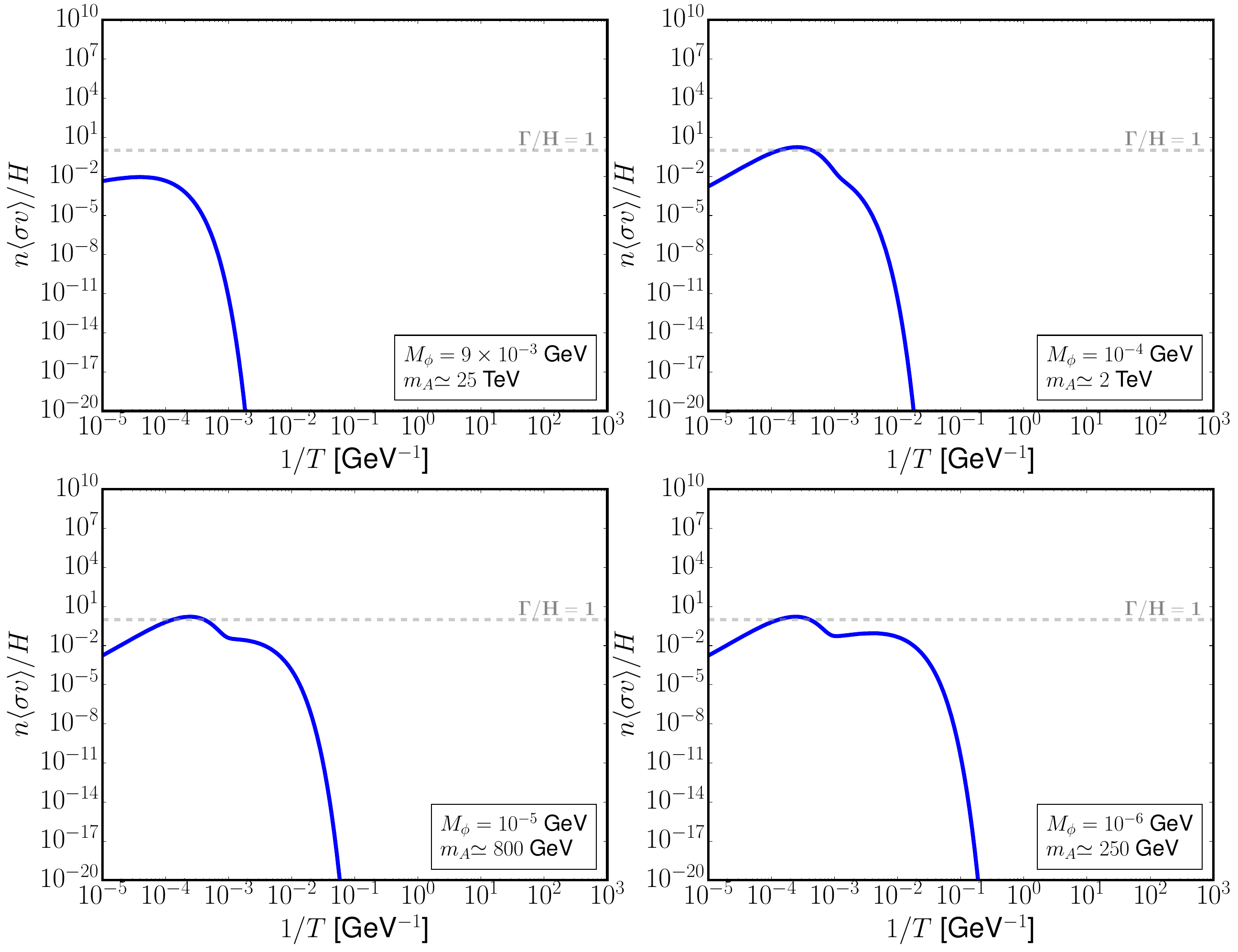}
\caption{Ratio between the interaction rates $\Gamma$ and the Hubble rate $H$ for $A$ varying $M_\phi$, and fixing $v_1 = 10^{10}$ GeV, $\lambda_2 = 10^{-12}$, and $\lambda_3 = 10^{-6}$. }
\label{Fig:Ms_thermal}
\end{figure*}

\begin{figure*}
\centering
\includegraphics[width=0.7\textwidth]{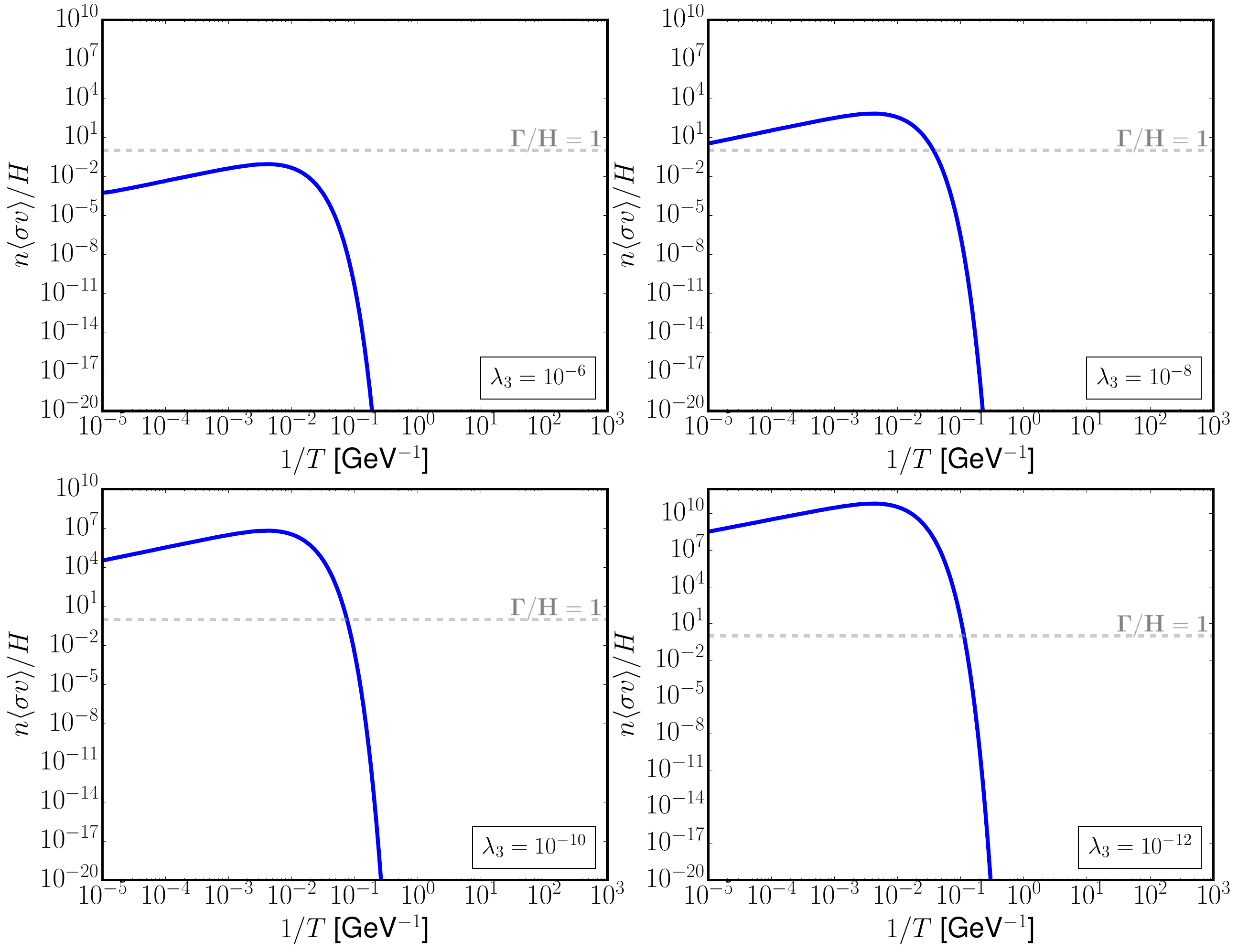}
\caption{Ratio between the interaction rates $\Gamma$ and the Hubble rate $H$ for $A$ varying with $\lambda_3$ and fixed $v_1=10^{12}$ GeV, $\lambda_2 = 10^{-12}$, $M_\phi = 10^{-8}$ GeV, and $m_A \simeq 250$ GeV. }
\label{Fig:lambda6_thermal}
\end{figure*}

Fig.~\ref{Fig:Ms_thermal} shows the behavior of $\Gamma/H$ fixing all the parameters and varying $M_\phi$. In the first upper panel, we display the result for the higher value of $M_\phi$. As we decrease it, the mass of $h^{\prime}$ increases and then the resonance of $h^{\prime}$ appears, as seen  in the last panel. Finally, when we fix all the parameters and vary $\lambda_3$, we obtain what is shown in Fig.~\ref{Fig:lambda6_thermal}. In this case, as  $\lambda_3$ decreases  the curves increases, which is expected since  the coupling between $h^\prime$ and SM particles is inversely proportional to $\lambda_3$ and $v_1$.

Finally, we can now compute the relic density in both scenarios with the viable parameter spaces that imply freeze-out or freeze-in. Additionally, it is important to note that Higgs can decay into inflaton through $h \to h^\prime h^\prime$. However, due to the extremely tiny values of parameters imposed by inflation, this branching ratio respects the bound of Higgs invisible decay. Note that we are particularly interested in computing only the relic abundance of $A$. This is so because  $A$ cannot be probed through direct detection experiment due the cancellation mechanism shown in \cite{Gross:2017dan}. Additionally, it's worth noting that one-loop processes, as discussed in \cite{Azevedo:2018exj}, can contribute to direct detection signal. However, it's important to mention that delving deeper into these one-loop processes falls outside the scope of this article.

\subsection{Freeze-in}

As shown in Fig.~\ref{Fig:freezeout_in}, there is a set of parameters that implies that $A$ is non-thermally produced. In this case, because   $g_{hAA}\,,\,g_{h^{\prime}AA}$ are very small, the DM is produced predominantly through inflaton decay in the early Universe \cite{Pallis:2004yy,Hall:2009bx,Chung:1998rq,Giudice:2000ex}. Initially, the energy density of the Universe is only composed of the energy density of inflaton $\rho_{h^\prime} = m_{h^\prime} n_{h^\prime}$, until it decays at the reheating temperature $T_R \sim 0.1 \sqrt{\Gamma_{h^{\prime}} m_{Pl}}$, where $\Gamma_{h^{\prime}}=\Gamma_{h^\prime\to hh}+\Gamma_{h^\prime \to A A}$ with
\begin{eqnarray} 
   && \Gamma_{h^\prime\to hh} \approx \frac{\lambda_3^2 v^2_1}{32\pi m_{h^{\prime}}}\,,\nonumber \\
   && \Gamma_{h^\prime \to A A} \approx  \frac{\lambda_2^2 v_1^2}{32\pi m_{h^{\prime}}}\,. 
   \label{decaywidth}
\end{eqnarray}
For the range of values of the parameters $\lambda_2\,,\,\lambda_3\,,\,v_1$  used here, we have that the channel $\Gamma_{h^\prime\to hh}$ is the dominating one\footnote{For $\lambda_3 =10^{-6}$, $\lambda_{2}=10^{-12}$ and $v_1=10^9$ GeV we have that $T_R\approx 10^8$ GeV.}. The inflaton decay into $N$ is kinematically forbidden ( $m_{h^\prime} \ll 2 m_N$).

Subsequently, the energy density of the Universe has additional contributions from  radiation $\rho_R$ and dark matter $\rho_{A} = m_{A} n_{A}$. In this period the Hubble rate is described by
\begin{equation}
    H^2 = \frac{8 \pi}{3 m_{Pl}^2} \left(\rho_{h^\prime} + \rho_{R} + \rho_{A} \right)\,.
\end{equation}

The evolution of the quantities $\rho_{h^\prime}$, $\rho_{A}$, and $\rho_R$ are given by the set of equations~\cite{Chung:1998rq},
\begin{eqnarray}
    \dot{n_{h^\prime}} + 3 H n_{h^\prime} + \Gamma_{h^\prime\to hh} n_{h^\prime} &=& 0 \nonumber \\ \label{eq:n_inf}\\
    \dot{n_{A}} + 3 H n_{A} + \langle \sigma v \rangle_{A A} \left(n_{A}^2 - n_{A}^{eq 2} \right) - \Gamma_{h^\prime\to A A} n_{h^\prime}  &=& 0 \nonumber \\ \label{eq:n_A}\\
    \dot{\rho_R} + 4 H \rho_R - \Gamma_{h^\prime\to hh}  n_{h^\prime} &=& 0 .\nonumber \\ \label{eq:rho_rad}
\end{eqnarray}

In order to solve numerically this set of coupled equations, we define the following parameters
\begin{equation}
    \Phi \equiv \frac{m_{h\prime} n_{h^\prime} a^3}{T_R} \,, \quad R   \equiv \rho_R a^4 \,, \quad X \equiv n_{A} a^3\,, \quad A \equiv a \times T_R\,.
\end{equation}

\begin{figure*}
    \centering
    \includegraphics[width=0.7\textwidth]{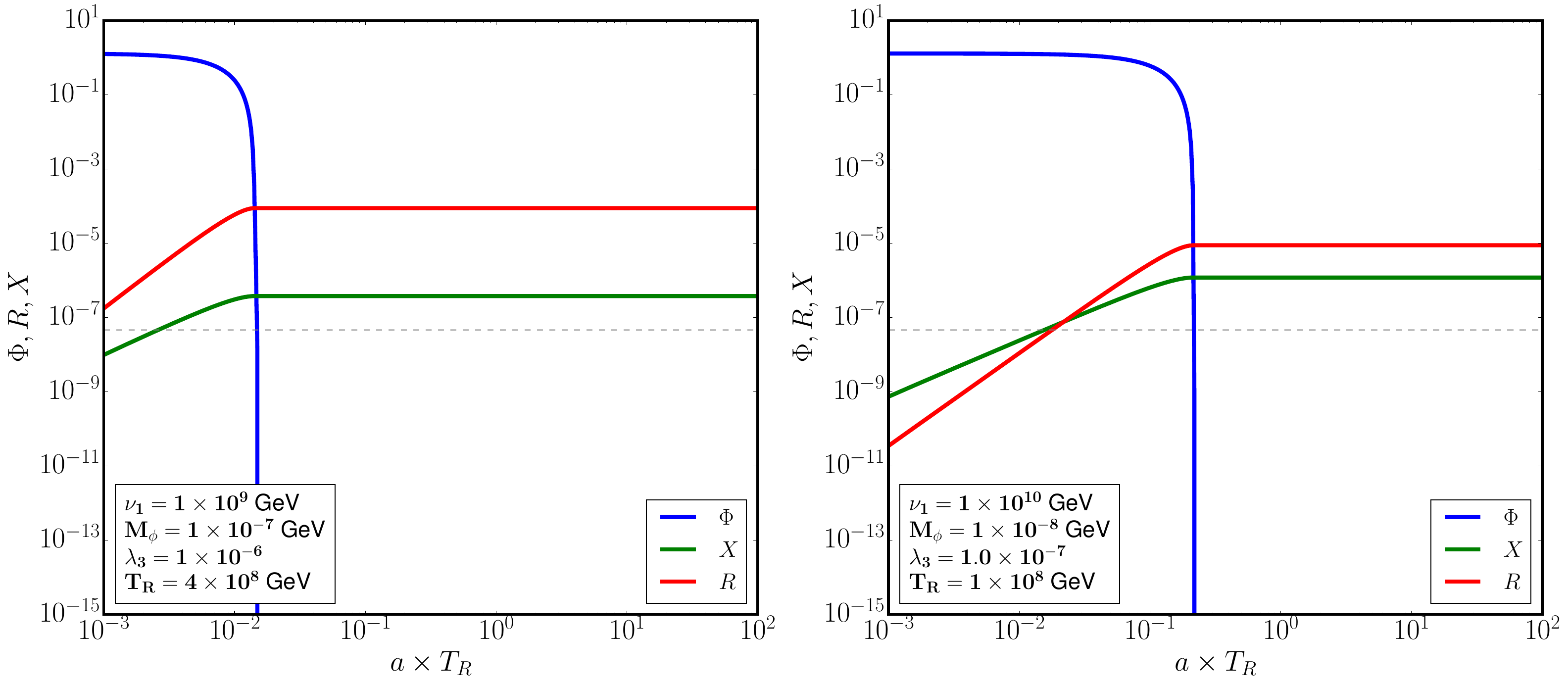}   
    \caption{Evolution of coupled Eqs.~\eqref{eq:n_inf},~\eqref{eq:n_A}, and~\eqref{eq:rho_rad}, where $\Phi$ (blue), $R$ (red) and $X$ (green), for different value of parameters. The grey dashed horizontal line represents the correct abundance.}
  \label{fig:Coupled_eq}
\end{figure*}

In Fig.~\ref{fig:Coupled_eq} we present the evolution of the set of Eqs.~\eqref{eq:n_inf},~\eqref{eq:n_A}, and~\eqref{eq:rho_rad} for different value of $\lambda_3$, $M_\phi$, and $v_1$. On the left panel of Fig.~\ref{fig:Coupled_eq} we
show the full solutions for $\Phi$ (blue curve) and $R$ (red curve), as well as the solution for
$X$ (green curve) for $M_\phi = 10^{-7}$ GeV, $\lambda_3 = 10^{-6}$, and $v_1 = 10^9$ GeV. On the right panel we use $M_\phi = 10^{-8}$ GeV, $\lambda_3 = 10^{-7}$, and $v_1 = 10^{10}$ GeV. As we can observe in both panels, initially the energy density of Universe is dominated by the inflaton (blue curve). The complete decay of $h^\prime$ corresponds in Fig.~\ref{fig:Coupled_eq} to the abrupt decrease of $\Phi$, which produces radiation (red curve) and dark matter (green curve) in the final state.

The estimation of relic abundance for this scenario can be given by \cite{Heurtier:2017nwl}
\begin{eqnarray}
    \Omega^0 h^2 &\simeq& 0.1\left(\frac{10^{13} \text{GeV}}{m_{h^\prime}} \right) \left(\frac{m_{A}}{\text{MeV}} \right) \left(\frac{\text{eV}}{T_0} \right) \left( \frac{T_R}{10^{10}\text{GeV}}\right) \nonumber \\
    &&\times Br(h^{\prime}\to AA)\,,
\end{eqnarray}
where $T_0 \sim 10^{-4}$ eV, such equation is in agreement with our result of set of coupled equations. Here we have $Br(h^{\prime}\to AA)\approx \frac{\lambda_2^2}{\lambda_3^2}=10^{-12}(\lambda_3=10^{-6})$.

\begin{figure*}
    \centering
    \includegraphics[width=0.7\textwidth]{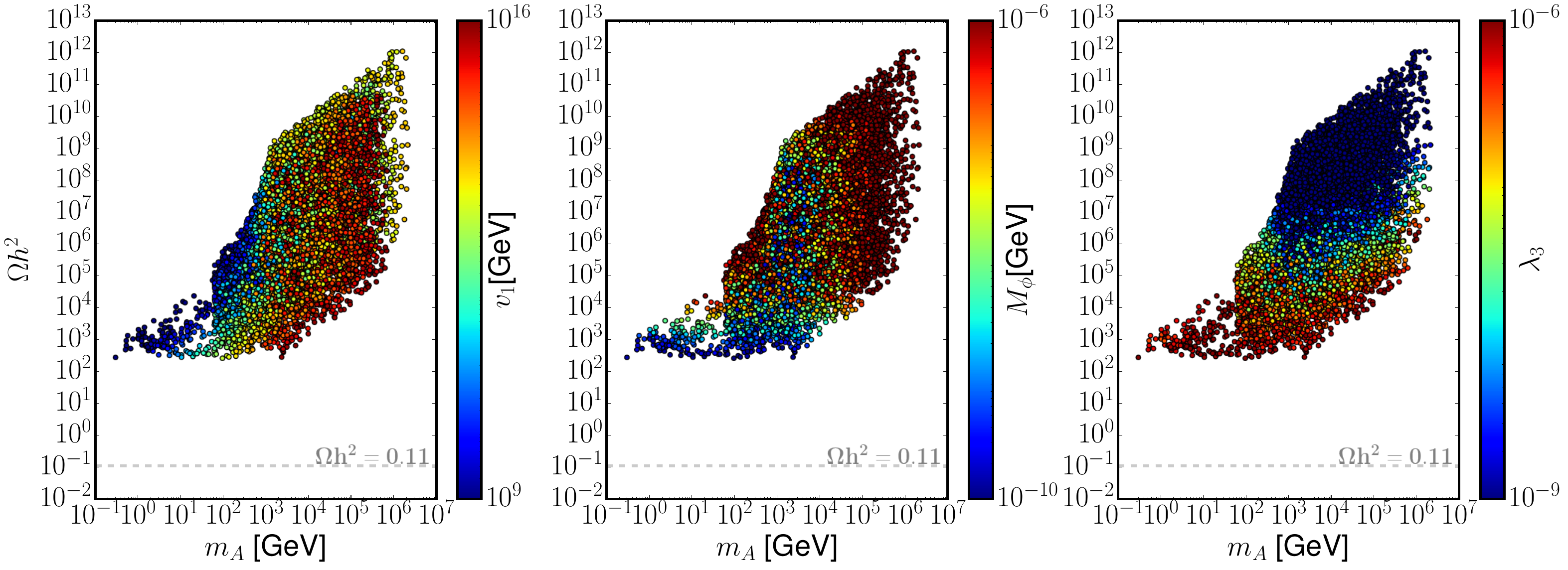}   
    \caption{The abundance of $A$ produced non-thermally through freeze-in in function of $m_A$. The color of points in the first plot represents the variation of $v_1$ as is indicated in the color bar, and the color of points in the second and third plots represents the variation of $M_\phi$ and $\lambda_3$. The gray horizontal dashed line represents the correct DM abundance.}
  \label{fig:freeze_in}
\end{figure*}

Additionally, in Fig.~\ref{fig:freeze_in} we compute the relic abundance for all parameter space that provides freeze-in scenario, which is shown in the right panel of Fig.~\ref{Fig:freezeout_in}. As we can see in Fig.~\ref{fig:freeze_in}, we can not obtain the correct DM relic abundance (represented by dashed horizontal gray line). Then we conclude here that  successful inflation is not compatible with freeze-in scenario.  

\subsection{Freeze-out}

In order to compute the relic abundance of $A$ for the freeze-out case, we need to solve the Boltzmann equation for the quantity $Y_A \equiv n_A/s$
\begin{equation}\label{eq:Boltz_freezeout}
    \frac{dY_A}{dx} = - \frac{x \langle \sigma v \rangle s}{H(m_A)} \left[ \left(Y_A \right)^2 - \left(Y_A^{eq} \right)^2 \right]\,,
\end{equation}
where $x\equiv m_A/T$, $T$ is the temperature, $s$ is the entropy density, $H(m_A)$ is the Hubble parameter in function of mass of $A$, the superior index ``$eq$'' stands for the distribution of equilibrium, and $\langle \sigma v \rangle$ is the thermally averaged cross section, described in Eq.~\eqref{thermally_avg}.

The relic abundance for this scenario is given by
\begin{equation}
    \frac{\Omega^0 h^2}{0.11} \simeq \frac{m_A}{100 \text{GeV}} \frac{Y^0}{4.34 \times 10^{-12}}\,,
\end{equation}
where the superior index ``$0$'' indicates quantities as measured today with $\Omega^0 h^2 \simeq 0.11$  given in Table. (\ref{tab:Cm})\footnote{Observe that $\omega_c \equiv \Omega^0 h^2 $}. The $Y^0$ can be obtained by integrating Eq.~\eqref{eq:Boltz_freezeout} from $x_0 = m_A/T^0$ to $x_\infty = m_A/T_\infty$, where $T_\infty \to \infty$, and $T^0$ is the temperature of the Universe today.

For this case, we compute the relic abundance of $A$ through MicrOmegas \cite{Belanger:2018ccd}, varying stochastically $\lambda_3$, $M_\phi$, and $v_1$, and additionally imposing that $\Gamma/H > 1$. The relevant annihilation processes that contribute to the abundance of $A$ are represented in Fig.~\ref{Fig:process}, with  lepton and fermion of SM in the final state.

\begin{figure*}
    \centering
    \includegraphics[width=0.8\textwidth]{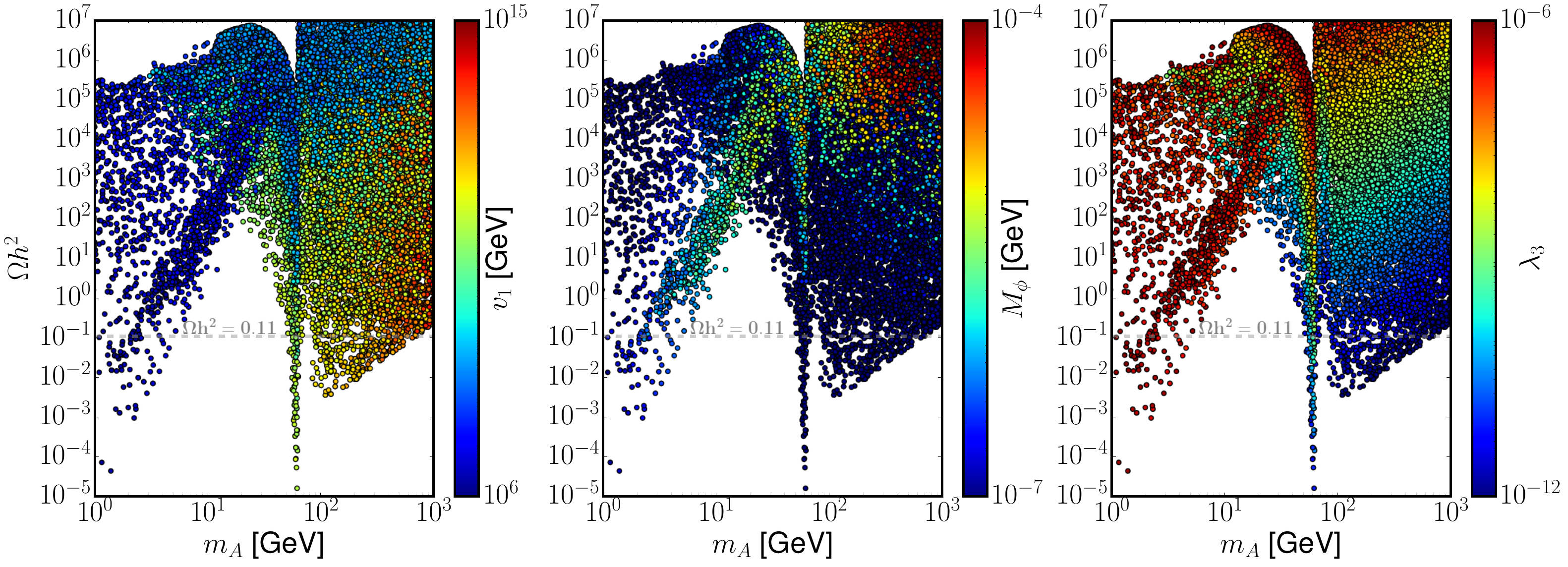}   
    \caption{The abundance of $A$ produced thermally through freeze-out varying stochastically $\lambda_3$, $M_\phi$, and $v_1$. The variation of $v_1$ is represented by the color of the points in the left panel. The variation of $M_\phi$ is represented by the color of the points in the right panel. The gray horizontal dashed line represents the correct DM abundance.}
  \label{fig:freeze-out}
\end{figure*}

The results of freeze-out scenario is shown in Fig.~\ref{fig:freeze-out}. There, we show the relic abundance $\Omega h^2$ as a function of $m_A$, where in the left panel the color of the points means the variation of $v_1$, whereas the middle and right panel present the variations with respect to $M_\phi$ and $\lambda_3$, respectively. Note that we have a suppression when $m_A = m_h/2$, which means the Higgs resonance. It's noteworthy to mention that the region of underabundance and correct abundance for lower dark matter mass is achieved when $h^\prime$ is lighter than the Standard Model Higgs boson. Additionally, we have noted that the correct abundance of DM can be obtained for $M_{\phi} \sim \left(10^{-8} - 10^{-6}\right)$ GeV and $v_1 \sim \left(10^{7} - 10^{12}\right)$ GeV.

\section{Conclusions}\label{sec:conclusion}

Inflation is the best paradigm describing the Universe's early stages, in which primordial fluctuations are identified as arising from quantum vacuum fluctuations stretched by an accelerated expansion. The cumulative effort employed to describe the initial conditions that lead to the pattern of cosmological perturbations resulted in many inflationary scenarios. In this work, we investigate the observational viability of a low-energy-inspired inflationary scenario called Inflection-Point Inflation (IPI) and propose a realization of such a model.

In particular, we use the Monte-Carlo sampling techniques associated with the Bayesian selection procedure to probe the IPI parameter space against current Planck data. We obtained constraints on the cosmological parameters that describe the evolution of the cosmological perturbations (Tab.~\ref{tab:Cm}), as well as the most restrictive constraints on the parameters of the underlying fundamental theory of inflation up to date, namely $M/m_p = 1.97^{+0.94}_{-0.70}$ for the amplitude of the inflaton field during horizon crossing moment, $\log_{10}\left(\lambda\right) = 11.82^{+0.46}_{-0.75}$ for the renormalized quartic coupling and its running $\log_{10}\left(|\beta_{\lambda}|\right) = 11.37^{+0.50}_{-0.82}$, both defined at the renormalization scale $M$.

Regarding the result for $\beta_\lambda$, we obtained negative values for the running of the quartic coupling $\beta_\lambda= - |\beta_{\lambda}|$, which indicates a non-negligible interaction between the inflaton field and fermions at the inflationary scale. This outcome is particularly interesting in the context of the type-I seesaw mechanism, where the inflaton's vacuum expectation value may generate Majorana masses for the active and sterile neutrinos \cite{Rodrigues:2020dod}, a possibility that will be explored in a forthcoming communication. We also investigated the consequences of the cosmological constraints in an underlying field theory where the inflaton composes a neutral complex scalar singlet with the dark matter. The inflaton is the real component, while the dark matter is the imaginary part. To recover the negative value of  $\beta_\lambda$, we added  a neutral fermion singlet to the standard model of particle physics. 

We used the parametric space required by successful inflation to investigate if such a model possesses a viable dark matter candidate. We then studied the case in which dark matter is thermally (freeze-out) and non-thermally produced (freeze-in), as the space of parameters does not favor any type of dark matter production. We found that successful inflation demands that dark matter be produced thermally.  It is also worth mentioning that this dark matter scenario escapes direct detection, being, in principle, possibly probed through indirect detection or at the LHC.

Finally, we highlight the applicability of the analysis carried out here and the predictability of the IPI scenario. The obtained constraints are easily translated to any realization of this scenario at lower energies, presenting an excellent alternative to probe new physics at energies within reach of future particle colliders.

\acknowledgments

JGR acknowledges financial support from the Programa de Capacita\c{c}\~ao Institucional do Observat\'orio Nacional - PCI-DA fellowship. RvM is suported by Funda\c{c}\~ao de Amparo \`a Pesquisa do Estado da Bahia (FAPESB) grant TO APP0039/2023. CASP  was supported by the Conselho Nacional de Desenvolvimento Cientifico e Tecnol\'ogico (CNPq) research grant No. 311936/2021-0. JSA is supported by CNPq grant No. 307683/2022-2 and Funda\c{c}\~ao de Amparo \`a Pesquisa do Estado do Rio de Janeiro (FAPERJ) grant No. 259610 (2021).

\appendix

\bibliography{apssamp}

\end{document}